\documentclass[12pt]{iopart}

\usepackage{graphicx}
\usepackage{iopams}

\newcommand{\PRA}{{\it Phys. Rev.} A }

\usepackage[usenames]{color}

\begin{document}

\title[]{Demonstrating mesoscopic superpositions in double-well Bose-Einstein condensates}

\author{T~J Haigh, A~J Ferris and M~K Olsen}

\address{ARC Centre of Excellence for Quantum-Atom Optics, 
School of Physical Sciences, University of Queensland, Brisbane, 
QLD 4072, Australia.}
\ead{tjhaigh@physics.uq.edu.au}
\begin{abstract}
The availability of Bose-Einstein condensates as mesoscopic or macroscopic quantum objects has aroused new interest in the possiblity of making and detecting coherent superpositions involving many atoms. In this article we show that it may be possible to generate such a superposition state in a reasonably short time using Feshbach resonances to tune the inter-atomic interactions in a double-well condensate.

We also consider the important problem of distinguishing whether a coherent superposition or a statistical mixture is generated by a given experimental procedure. We find that unambiguously distinguishing even a perfect `NOON' state from a statistical mixture using standard detection methods will present experimental difficulties.

\end{abstract}

\section{Introduction}
The idea of a macroscopic state demonstrating quantum mechanical behaviour was introduced by Schr\"{o}dinger in 1935~\cite{schrodingerscat}. His famous thought experiment considered how a macroscopic entity (in this case a domestic cat) could evolve into a superposition of two distinct physical states (alive and dead) when entangled with a microscopic system that obeyed the laws of quantum mechanics. This possibility of a macroscopic system being simultaneously in two distinct physical states was initially considered to be a flaw in quantum mechanical theory~\cite{schrodingerscat}. However, experiments have shown the predictions of quantum mechanics to be correct as superpositions of macroscopically distinct physical states have been produced in a variety of systems. These are often referred to as `Schr\"{o}dinger cat states'. For example, a group of six beryllium ions has been put into a superposition of two hyperfine states~\cite{beryllium}, and persistent currents (of a few $\mu$A) of opposing circulation in SQUIDs have been detected~\cite{SQUIDs}. 

Bose-Einstein condensates (BECs) are attractive candidates for generating macroscopic superpositions due to the large number of atoms that share a single quantum state. This may provide a useful system in which to further test the validity or boundaries of the assumption of macroscopic realism. A macroscopic superposition has yet to be demonstrated in a Bose-Einstein condensate, although there exist numerous proposals for generating either a superposition of relative phase or number states~\cite{2compCirac1997,Dunningham,Zoller03,GordonSavage,phasesupn1,phasesupn2}. In the following we consider a particular kind of superposition state of a single component BEC in a double well, that is, a superposition of the two states where the entire condensate is localised in one of the wells (sometimes referred to as a `NOON' state). If realisable, this kind of superposition promises to be useful in quantum information applications and precision interferometry, due to the measurement uncertainty scaling inversely with the number of particles (the so-called Heisenberg limit).

A major difficulty in realising macroscopic quantum superpositions is decoherence, which occurs when interactions with the surrounding environment cause the pure superposition state to decay into a statistical mixture. However, this paper is not concerned with avoiding decoherence in the realisation of macroscopic superpositions. Instead, we concentrate on measurements aimed at distinguishing a coherent superposition from a statistical mixture. We find even in a decoherence-free environment, demonstrating a superposition presents several practical challenges. Once realised, measurements of the purity of the state could be useful in studying, for example, rates of decoherence.

This paper begins with an introduction of the two-mode description of a double-well condensate in Section~\ref{sec2}. In Section~\ref{sec3} we consider measurements aimed at distinguishing between a coherent superposition and a statistical mixture. We focus on quadrature-based measurements, analogous to those used in quantum optics, that can be realized with a Ramsey-type interference experiment. The coherence of the NOON state is evident in parity measurements of the number distribution after the two modes are interfered --- a difficult measurement with standard atom counting techniques. We analyse the effects of atomic interactions and discuss atom loss during the interference procedure. Finally, we show in Section~\ref{sec4} that a mesoscopic superposition of 20 atoms could be generated in a reasonable time frame if a Feshbach resonance can be used to tune the atomic interactions, before concluding in Section~\ref{sec5}.

\section{Theoretical Models}

\label{sec2}

\subsection{Two-mode approximation}

The Hamiltonian for a condensate in an external trapping potential, $V_{ext}(\vec{r})$, is
\begin{equation}
\hat{\mathcal{H}}=\int{ d \vec{r} \left[ \frac{\hbar^2}{2m} \nabla \hat{\psi}^{\dagger} \cdot \nabla \hat{\psi} +V_{ext}(\vec{r}) + \frac{\hbar U_0}{2}\hat{\psi}^{\dagger} \hat{\psi}^{\dagger} \hat{\psi} \hat{\psi}   \right] } , 
\label{hamiltonian}
\end{equation}
where $\hat{\psi}$ is the field operator for the condensate, and the non-linear interaction parameter is $U_0=4\pi a \hbar / m$, ($a$ is the s-wave scattering length describing two-body collisions within the condensate, $m$ is the atomic mass). We consider the case where the external potential provides a double well confinement for the condensate. Double well potentials can be generated by an optical lattice with an additional harmonic confinement to reduce the number of occupied lattice sites to two~\cite{dwsqueezing}. They can also be realised on chips, where suitably arranged current carrying wires create a magnetic confinement for the condensate atoms~\cite{dwinterferometrychip}. 

When a double well potential is considered, the above Hamiltonian can be simplified by making use of the two-mode approximation. This means we consider each atom to be in some linear superposition of being in the left well and being in the right well.

We consider the zero temperature case, where all atoms in the system are condensed. If the ground state energies of the condensate in the two single (and separate) wells are sufficiently separated from the energies of the condensate in all other excited single particle states, transitions to or from the two modes of interest and these higher lying states can be neglected. This is required for the two-mode description to be valid. In the two-mode approximation, the field operator is expanded as
\begin{equation}
\hat{\psi}(\vec{r})\approx\phi_L(\vec{r}) \hat{a}_L + \phi_R (\vec{r})\hat{a}_R ,
\end{equation}
where $\hat{a}_L$ and $\hat{a}_R$ are discrete Bose annihilation operators for the left and right well respectively, and $\phi_{L/R}$ are the ground state spatial wave functions of the condensate in the left and right wells.

Substituting this into equation (\ref{hamiltonian}), we find an effective Hamiltonian
\begin{eqnarray}
\hat{\mathcal{H}}_{eff}&=&\hbar E_L \hat{a}_L^\dagger \hat{a}_L+ \hbar E_R \hat{a}_R^\dagger \hat{a}_R + \frac{\hbar U_L}{2} \hat{a}_L^\dagger \hat{a}_L^\dagger \hat{a}_L \hat{a}_L \\ \nonumber
&&+\frac{\hbar U_R}{2} \hat{a}_R^\dagger \hat{a}_R^\dagger \hat{a}_R \hat{a}_R
-\hbar \kappa\left(\hat{a}_L^\dagger \hat{a}_R+\hat{a}_R^\dagger \hat{a}_L  \right),
\label{effhamiltonian}
\end{eqnarray}
where we have neglected the spatial overlap of the left and right well densities. The single well bound state energies, $E_{L/R}$, are
\begin{equation}
E_{L/R}=\frac{1}{\hbar}\int{d\vec{r}~ \phi_{L/R}^*(\vec{r}) \left(\frac{-\hbar^2}{2m}\nabla^2 + V_{ext}(\vec{r}) \right) \phi_{L/R}(\vec{r}) } .
\end{equation}
$\kappa$, the tunnel coupling, is
\begin{equation}
\kappa=\frac{-1}{\hbar}\int{d\vec{r}~ \phi_{L/R}^*(\vec{r}) \left(\frac{-\hbar^2}{2m}\nabla^2 + V_{ext}(\vec{r}) \right) \phi_{R/L}(\vec{r}) } ,
\end{equation}
and the effective non-linear interaction terms are
\begin{equation}
U_{L/R}=U_0 \int{d\vec{r}~ \vert \phi_{L/R}(\vec{r}) \vert^4 } .
\label{interactioneqn}
\end{equation}
For the remainder of this paper we assume a symmetric potential, where $E_L=E_R=0$ and $U_L=U_R \equiv U$.

\subsection{Fixed number representation}

By ignoring all possibility of atom loss and decoherence, we can efficiently represent the $N$-body wave function using the basis $\vert N-n,n \rangle$, representing states with $N-n$ atoms in the left well, and $n$ atoms in the right well. Any wave function can be written as a superposition of these number states, i.e.
\begin{equation}
\vert \psi(t)\rangle = \sum_{n=0}^{N}{c_n(t) \vert N-n,n \rangle},
\end{equation}
where $\sum_n \vert c_n (t)\vert^2=1$. In this representation, the expectation value of the number of atoms in the left well is
\begin{equation}
\langle N_{L} \rangle = \sum_n{\left(N-n \right) \vert c_n(t) \vert^2}  ,
\end{equation}
and the variance in the number difference is
\begin{equation}
V = \sum_n{(N-2n)^2 \vert c_n \vert^2}-\left[\sum_n{(N-2n) \vert c_n \vert^2}  \right]^2 .
\end{equation}
For any initial state, $\vert \psi \rangle$, these coefficients have a time-dependence given by
\begin{equation}
c_n(t) = \langle N-n,n \vert \psi(t)\rangle = \langle N-n,n \vert e^{-i\bar{\mathcal{H}}t/\hbar} \vert \psi(0)\rangle .
\end{equation}
Taking the derivative with respect to time, we find~\cite{TindleWalls}
\begin{equation}
\frac{d c_n(t)}{dt}=\frac{-i}{\hbar} \langle N-n,n \vert \bar{\mathcal{H}} \vert \psi(t) \rangle  .
\end{equation}
Inserting the Hamiltonian, (\ref{hamiltonian}), into this expression gives the equations of motion for the number state coefficients
\begin{eqnarray}
 i\hbar \frac{dc_n(t)}{dt} &  = & \hbar U_L\left[\left(N-n\right)^2-\left(N-n\right)  \right] c_n(t)
 +\hbar U_R \left[n^2-n    \right]  c_n(t) \nonumber \\
&-& \hbar\kappa \sqrt{\left(N-n+1 \right) n } ~c_{n+1}(t)  -\hbar\kappa \sqrt{\left(n+1 \right) \left(N-n \right)} ~c_{n-1}(t)  .
\label{cnbydt}
\end{eqnarray}
The ground states and dynamics of a condensate in the double well can then be found by solving these equations numerically.

\begin{figure}[t]
\begin{center}
\includegraphics[width=0.32\columnwidth]{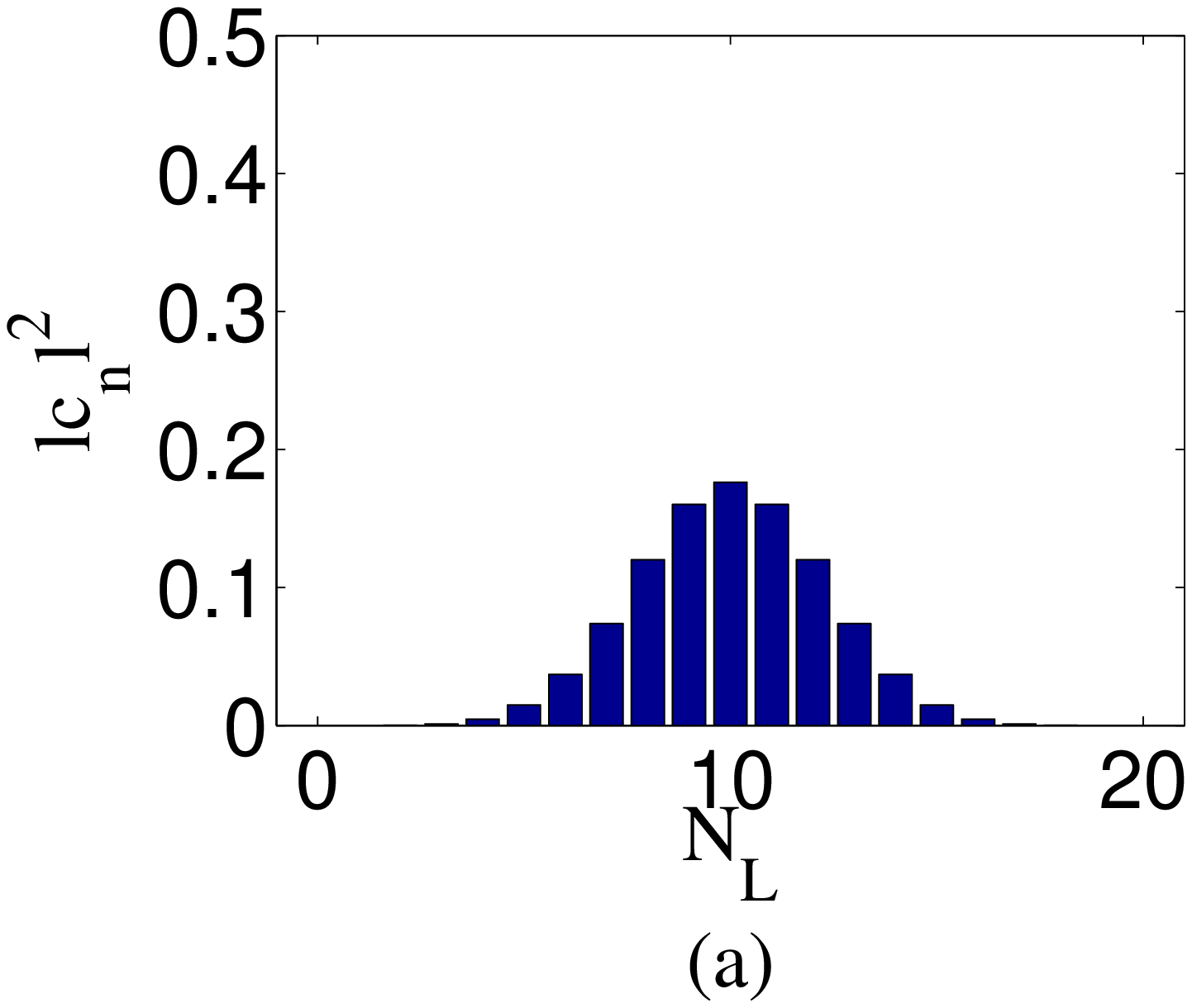}
\includegraphics[width=0.32\columnwidth]{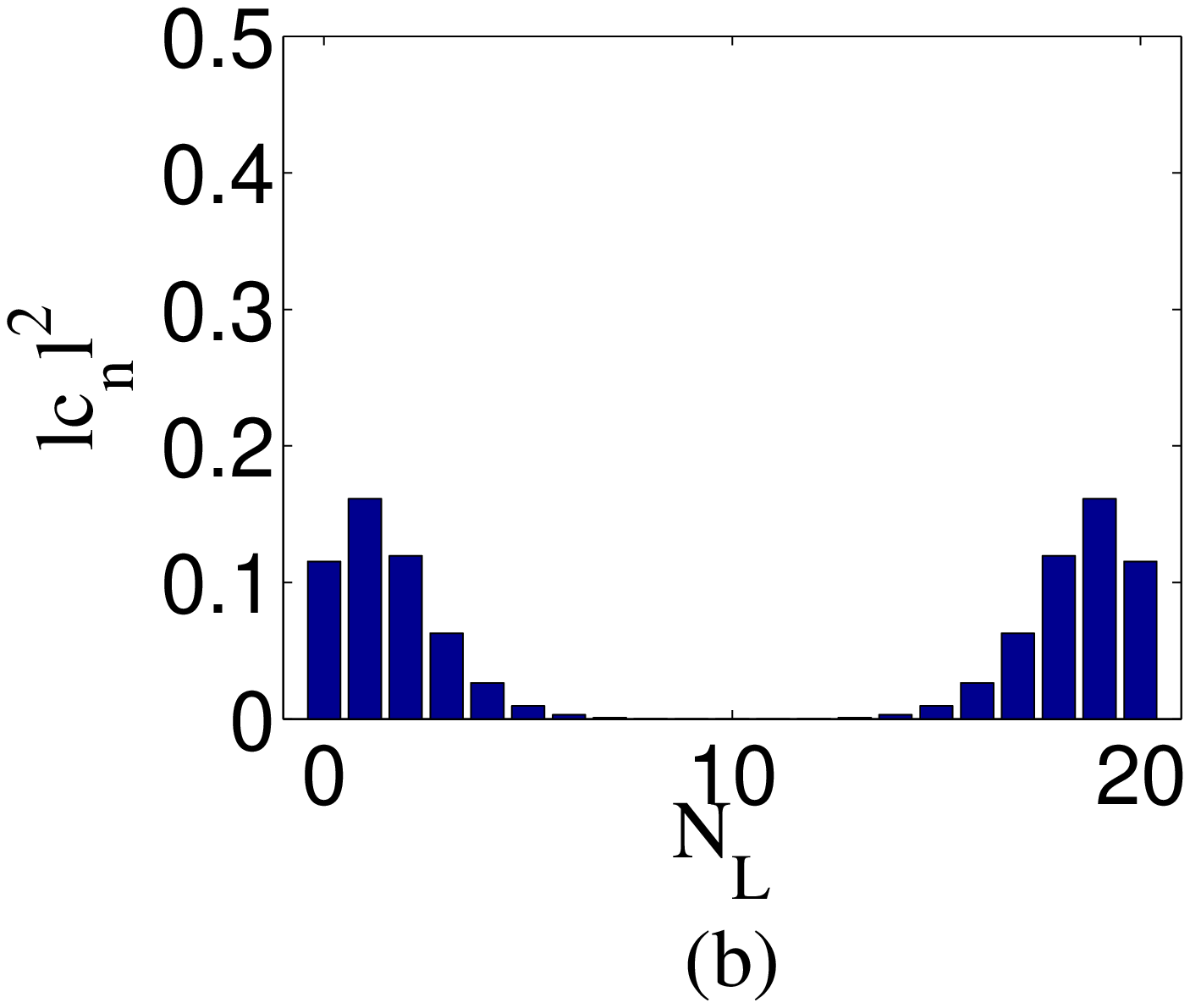}
\includegraphics[width=0.32\columnwidth]{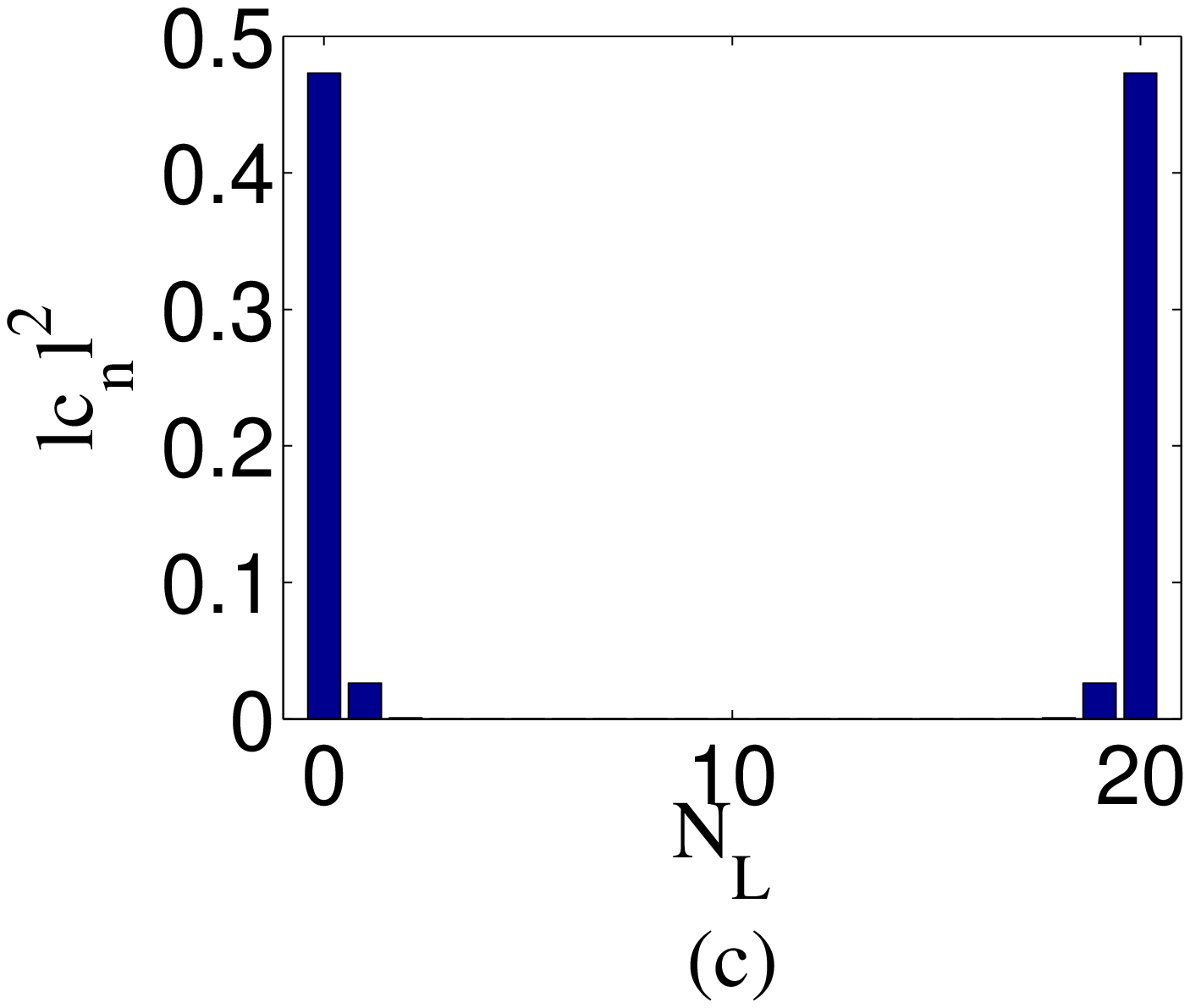}
\end{center}
\caption{Ground state probability distributions for weak, intermediate, and strong attractive interactions. $N_L$ is the atom number in the left well. Interaction strengths are (a) $U/\kappa=0$, (b) $-0.1$, and (c) $-0.5$.}
\label{fig:gndstates}
\end{figure}

Fig.~\ref{fig:gndstates} shows the ground state probability distributions for the atom number in the left well (found using imaginary time propagation~\cite{ITprop}) for three different regimes - negligible, intermediate, and strongly attractive interactions. The variance in the number difference for these are 12, 293, and 396 respectively (a perfect NOON state would have a variance of 400).

When non-linear interactions are weak compared to the tunneling, the ground state number distribution is essentially binomial (Fig.~\ref{fig:gndstates} (a)). For large N this can be approximated as a Poissonian distribution and the ground state would be a coherent state (which has equal mean and variance of atom number in each well). For intermediate attractive interactions, ground states exist consisting of two well separated peaks in the number state distribution. These can be considered as superpositions of distinct physical states, and we refer to them as `mesoscopic' superpositions~\cite{Reid}.

The ground state in the limit of infinitely strong attractive interactions is a macroscopic superposition of the entire condensate localized in each well. This can be understood by realizing that it is energetically preferable for the condensate to be localized in one well, but that given the symmetry of the double well potential these two localized states are degenerate. Thus in the ground state it is equally likely that the condensate will be found entirely in one well as in the other (hence the two peaks in the probability distribution). This state is the macroscopic superposition that we are interested in. Fig.~\ref{fig:gndstates} (c) shows that with a ratio of non-linear interaction strength to tunnelling rate $U/\kappa = -0.5$, the ground state of the double well condensate is close to, but not quite, an ideal superposition. We return to the problem of creating such a state in Sec.~\ref{sec4}.

\section{Verifying a superposition state}

\label{sec3}

We now consider the problem of demonstrating that a state is a macroscopic quantum superposition, as opposed to a statistical mixture. This will allow tests of macroscopic realism, may be an important tool for studying decoherence, and is a prerequisite to building a practical interferometer using NOON states. 

The obvious measurement to make on such a system is the atom number in each well. For the NOON state, we have equal probability of finding $N$ atoms in the left well and none in the right, or vice-versa. However, such a measurement can not distinguish the coherent NOON state 
\begin{equation}
 |\phi\rangle = \left(|N,0\rangle + e^{i\phi}|0,N\rangle\right)/\sqrt{2} \label{pure_state}
\end{equation}
from the statistic mixture with density operator
\begin{equation}
 \hat{\rho} = \left( |N,0\rangle \langle N,0| + |0,N\rangle \langle 0,N| \right) / 2. \label{mixed_state}
\end{equation}

There exist complimentary measurements that distinguish between a coherent superposition and a statistical mixture. One method of finding such an measurement is motivated by the expansion of the density operator of the statistical mixture as the average of all coherent phases $\phi$,
\begin{equation}
 \hat{\rho} = \frac{1}{4\pi} \int \left(|N,0\rangle + e^{i\phi}|0,N\rangle\right)\left(\langle N,0| + e^{-i\phi}\langle 0,N|\right) d\phi.
\end{equation}
We see that the determination of $\phi$ constitutes proof of coherence --- as $\phi$ is undefined for the mixed state.

Consider an arbitrary operator $\hat{x}$; for the mixed state in (\ref{mixed_state}), the expectation value of $\hat{x}$ is the average of the two pure, separable states,
\begin{equation}
\langle \hat{x} \rangle_{mixed} = \frac{1}{2}\left( \langle N,0|\hat{x}|N,0\rangle + \langle 0,N|\hat{x}|0,N\rangle \right).
\end{equation}
On the other hand, for the pure NOON state $|\phi\rangle$, the expectation value is
\begin{equation}
\langle \hat{x} \rangle_{pure} = \langle \hat{x} \rangle_{mixed} + \frac{1}{2} \left( e^{i\phi} \langle N,0|\hat{x}|0,N\rangle + e^{-i\phi} \langle 0,N|\hat{x}|N,0\rangle \right). \label{expectation_pure}
\end{equation}
The additional interference terms clearly display a dependence on the value of $\phi$. Therefore, an appropriate observable $\hat{x}$ has non-zero $\langle 0,N|\hat{x}|N,0\rangle$. An example of an operator that achieves this is $\hat{a}_L^{\dag N} \hat{a}_R^{N} + \hat{a}_R^{\dag N} \hat{a}_L^{N}$, which coherently transfers $N$ atoms from one well to the other or vice-versa. Unfortunately, there is no clear way of directly measuring this observable in an experimental setting. 

\subsection{Quadrature phase measurements}
\label{sec:quadrature}

We now consider a quadrature-based method for distinguishing the entangled NOON state from the statistical mixture. In this paper, we define the quadrature operator as
\begin{equation}
\hat{\mathcal{X}}_{\theta}=\hat{a}_L^{\dagger} \hat{a}_R e^{-i\theta}+\hat{a}_R^{\dagger}\hat{a}_L e^{i\theta}.
\label{quadoperator}
\end{equation}
The measurement of this observable can be achieved using simple linear interference and number measurements. Such a procedure is analogous to quantum optics experiments using a 50-50 beam splitter to interfere two photonic modes before intensity measurement, and allows access to phase information. Unlike common quantum optics experiments, both modes contain a similar number of atoms and neither mode can be interpreted as a local oscillator. This explains the difference between the above definition and the standard quadrature arising from homodyne measurement, proportional to $\hat{a}_Le^{i\theta}+\hat{a}_L^{\dagger}e^{-i\theta}$. Both definitions have been employed in theoretical discussions of BECs in the past~\cite{EPRquad,Ferris2008,Ferris2009}. 

To realize these quadrature measurements, we propose a Ramsey-type experiment (see Ref.~\cite{Ramsey} for a description of the Ramsey technique). Similar experiments have been proposed for double well condensates (for example, to detect a weak force~\cite{Corney}). The first step, after creating the superposition state, is to set the tunneling rate between the wells $\kappa$ and interaction strength $U$ to zero and let the system evolve for some time, $\delta t$, during which an energy imbalance exists between the wells (i.e. $E_L-E_R=\delta E$ is non-zero). The quadrature angle is set by $\delta E \delta t = \theta$. The second stage is to restore the symmetry of the wells (i.e. set $\delta E$ to zero), and switch on tunneling for a time of $\pi/ 4\kappa$ (this is analogous to a beam splitting operation). After this we make a measurement of the atom number in each well. The difference in atom number is exactly proportional to $\hat{\mathcal{X}}_{\theta}$. We simulate the entire procedure using the equations of motion of the number state coefficients (\ref{cnbydt}). It is then straightforward to numerically calculate the distribution of measurement outcomes, as given by $|c_n|^2$, and thus any moment (e.g. mean, variance, etc) of the the quadrature $\hat{\mathcal{X}}_{\theta}$.

\begin{figure}[t]
\includegraphics[width=0.32\columnwidth]{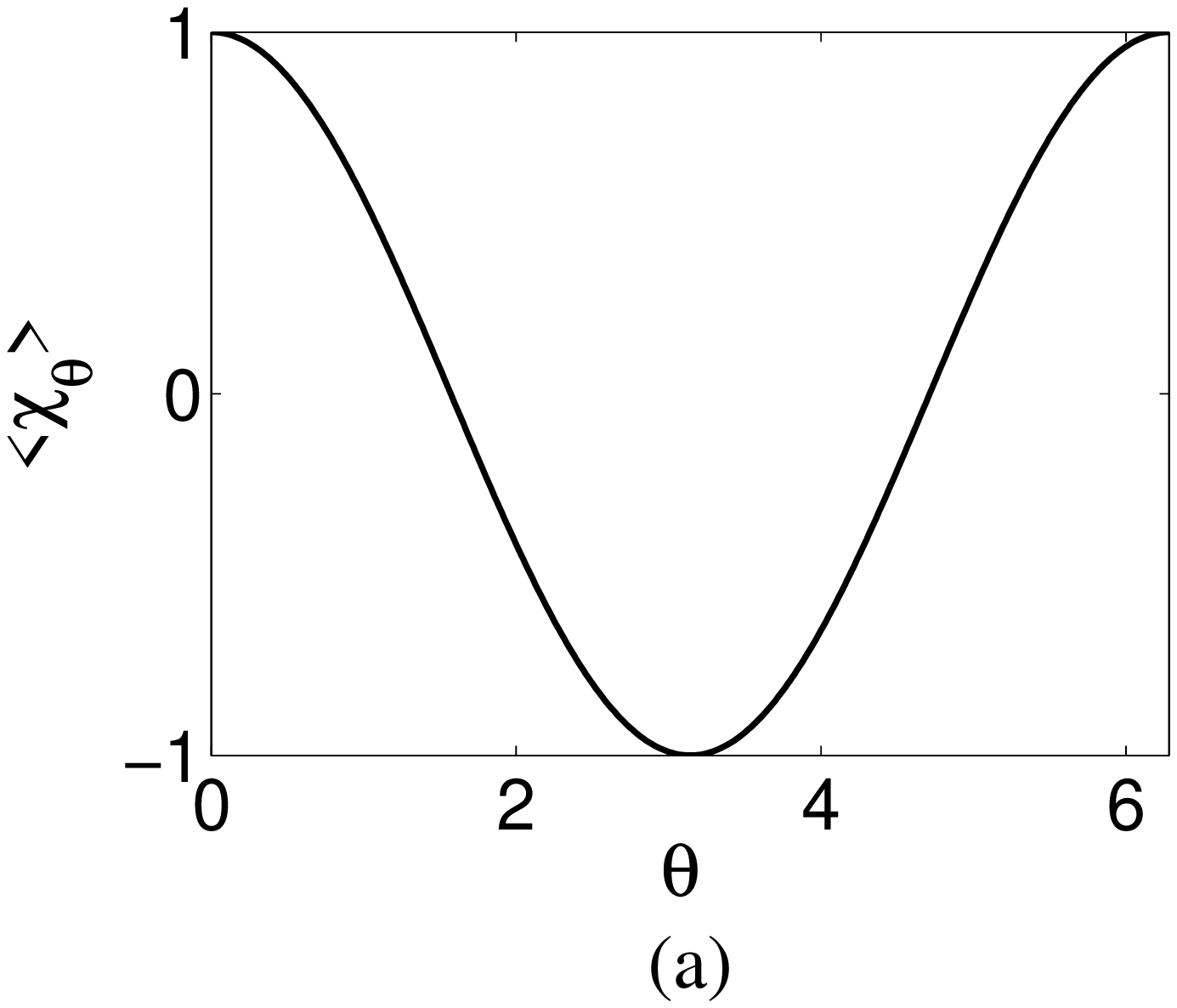}
\includegraphics[width=0.32\columnwidth]{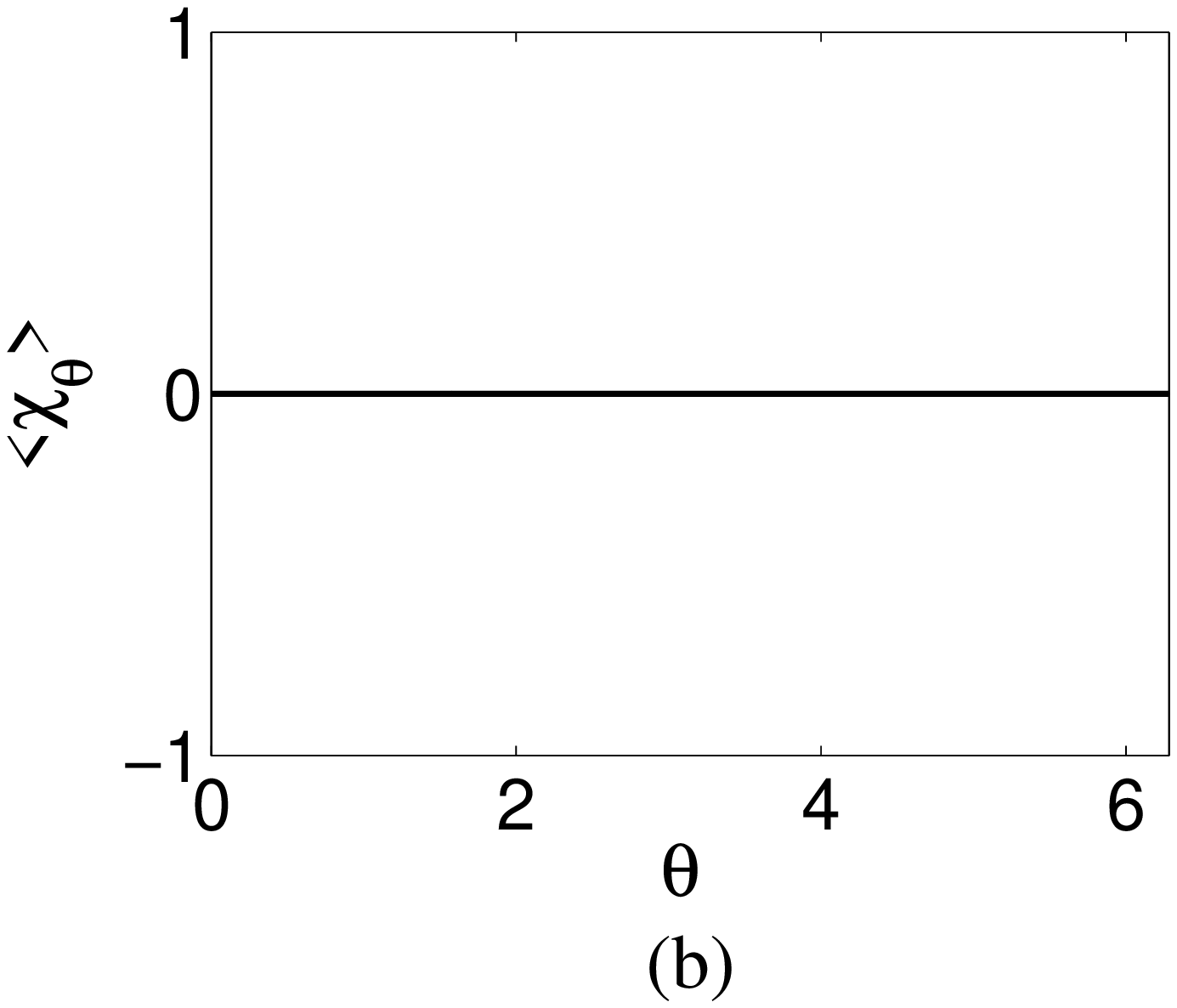}
\includegraphics[width=0.32\columnwidth]{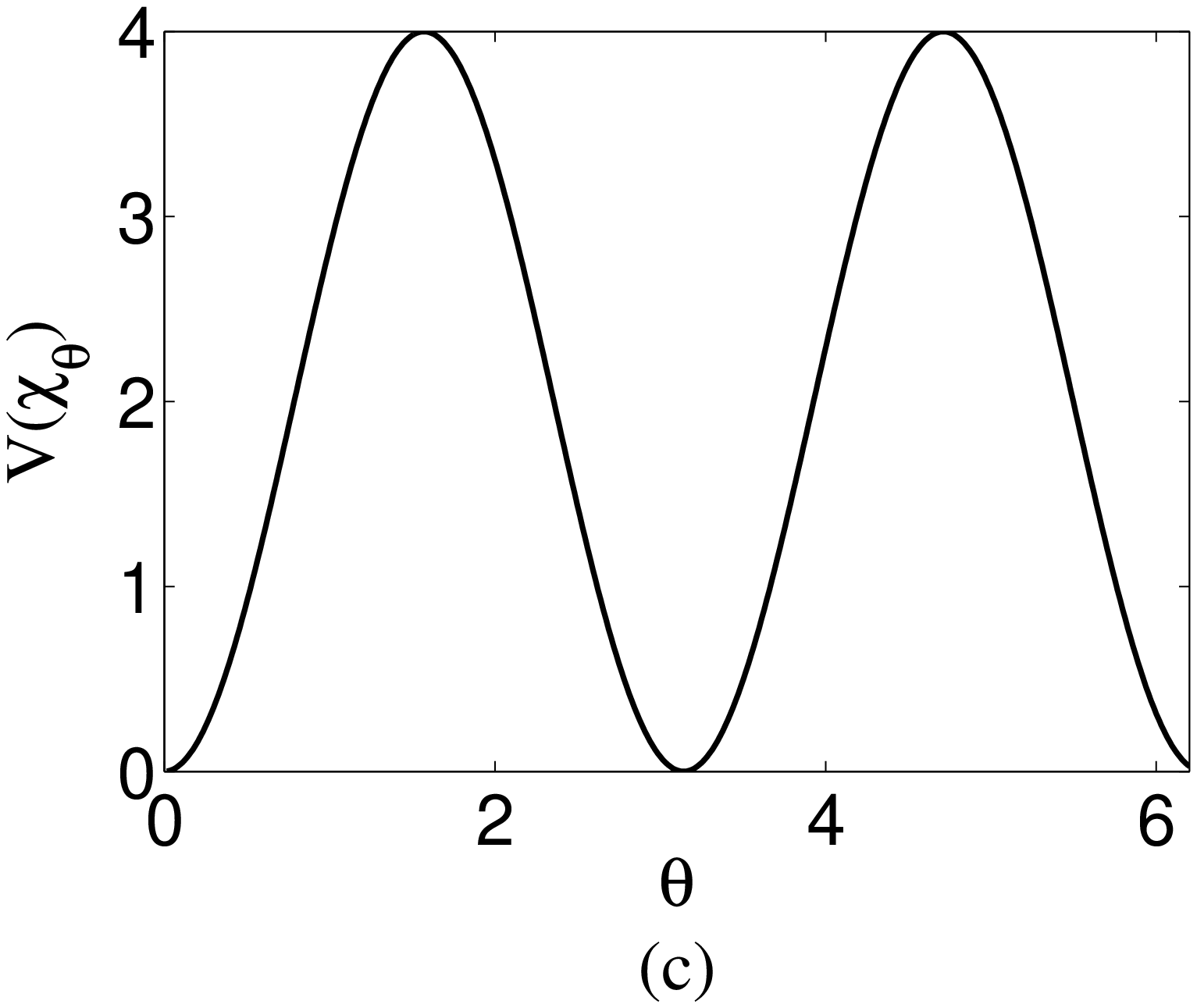}
\caption{(a) Interference fringes of $\langle \hat{\mathcal{X}}_{\theta} \rangle$ as a function of the accumulated phase shift for an ideal superposition containing 1 atom. (b) The same quantity for an ideal superposition containing 2 atoms displays no interference fringes. (c) For the two atom superposition state, the variance ($V(\hat{\mathcal{X}}_{\theta})=\langle \hat{\mathcal{X}}_{\theta}^2 \rangle-\langle \hat{\mathcal{X}}_{\theta} \rangle^2$) is sensitive to the phase $\theta$ accumulated during the Ramsey simulation.}
\label{Ramsey1atomUzero}
\end{figure}

The next step is to extract information about the off-diagonal terms of the density matrix by measuring the interference terms in (\ref{expectation_pure}). It is straightforward to see that $\langle N,0| \hat{\mathcal{X}}_{\theta} |0,N\rangle = 0$ for $N \ge 2$. In the special case of $N=1$, interference is observed by a sinusoidal dependence on the value of $\theta - \phi$, as seen in Fig.~\ref{Ramsey1atomUzero} (a) (where $\phi$ is the phase angle used in (13)). In Fig.~\ref{Ramsey1atomUzero} (b) we plot the same quantity for $N=2$, and it is unsurprising that inteference is lacking for this case.

However, higher moments of the quadrature measurements do contain information that can distinguish a pure NOON state from a classical mixture. For $N$ atoms, the $N$th moment $\langle \hat{\mathcal{X}}_{\theta}^N \rangle$ contains exactly one subterm equal to $\hat{a}_L^{\dag N} \hat{a}_R^{N} + \hat{a}_R^{\dag N} \hat{a}_L^{N}$, while all lower moments lack such a term. In Fig.~\ref{Ramsey1atomUzero} (c), we see interference fringes in the quadrature variance for the case $N=2$. The frequency of these fringes is doubled compared to the case $N=1$. 

In general the $N$ atom NOON state will display fringes with frequency $N$ times greater than the first-order coherence measured by $\langle \hat{\mathcal{X}}_{\theta} \rangle$ allows. Specifically, the $N$th quadrature moment contains terms proportional to $\cos\bigl(N(\theta-\phi)\bigr)$. It is this scaling that makes NOON states of interest for precision interferometry --- a NOON state with known $\phi$ could potentially be used to measure $\delta E \delta t$ with accuracy proportional to $N^{-1}$ (the so-called Heisenberg limit), compared with the $N^{-1/2}$ scaling typical when using `classical' inteferemetric techniques. Such scaling has been observed in single-photon experiments~\cite{Higgins}, but to-date neither with atoms nor NOON states.

\begin{figure}[t]
\begin{center}
\includegraphics[width=0.32\columnwidth]{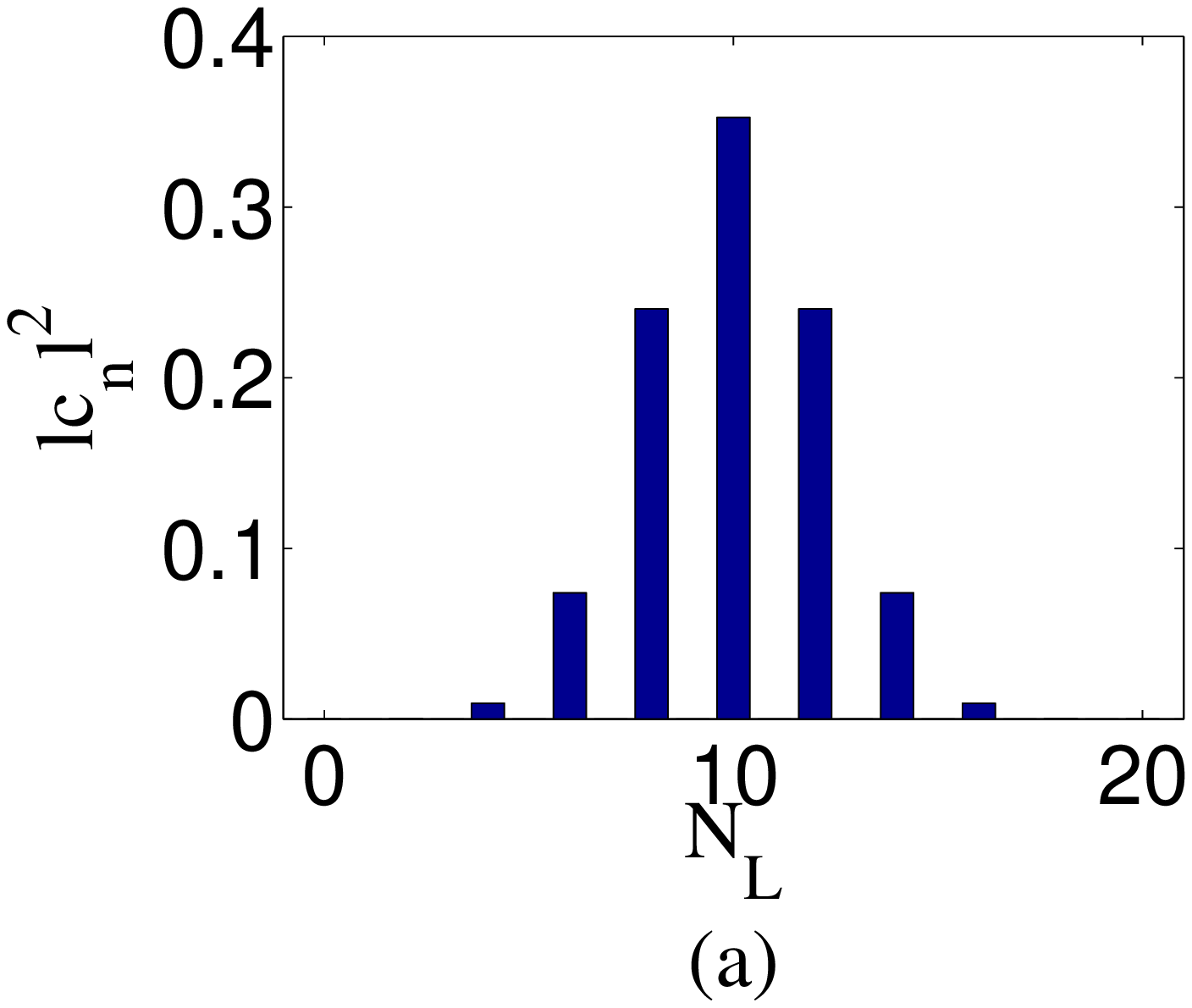}
\includegraphics[width=0.33\columnwidth]{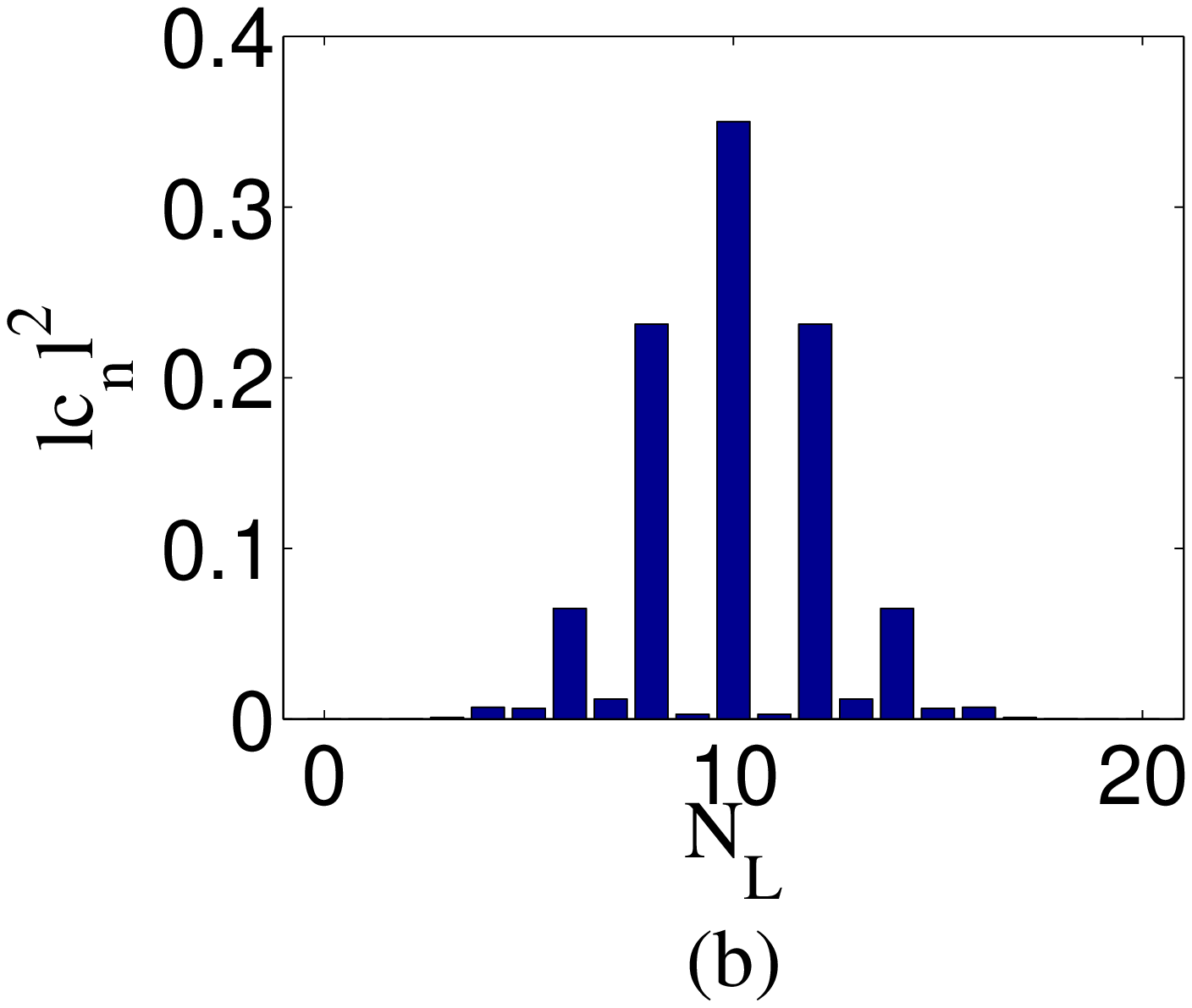}
\includegraphics[width=0.32\columnwidth]{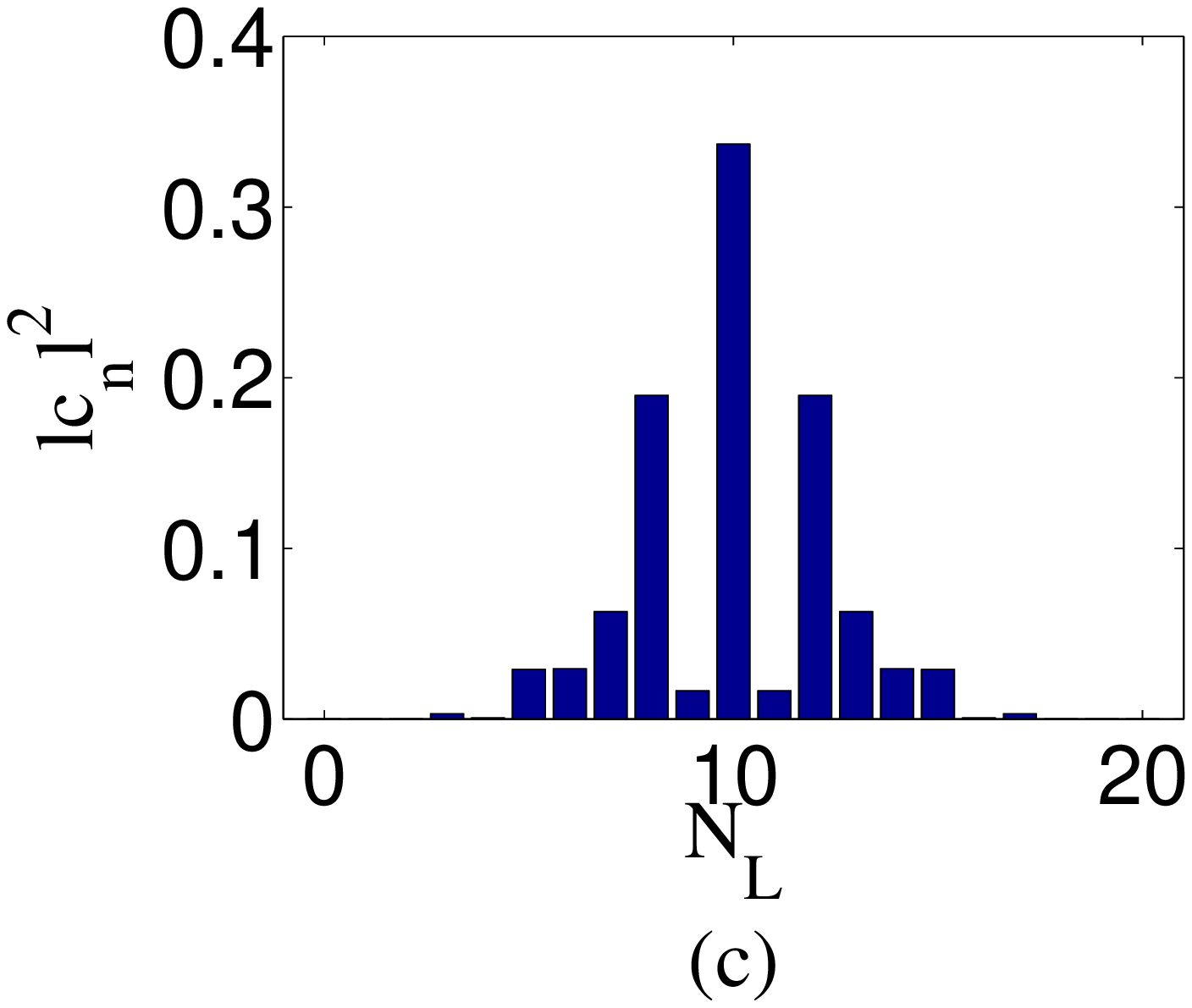}
\caption{Atom number distribution $|c_n|^2$ after the Ramsey procedure is simulated, with accumulated phase $\pi/2$. Initial state has 20 atoms and $\phi=0$. Non-linear strength $U/\kappa$ is (a) zero, (b) $-0.01$, and (c) $-0.025$.}
\label{ramseycninterference}
\end{center}
\end{figure}

The quadrature moments are intrinsically linked with the number distribution $|c_n|^2$ after the Ramsey interference procedure, and so it follows that the off-diagonal terms in Eq.~(\ref{expectation_pure}) are directly visible in this distribution. We have plotted the output in Fig.~\ref{ramseycninterference} (a) for $\theta = \pi/2$, and we observe a pattern where each second $\vert c_n\vert^2$ is zero. This interference pattern is sensitive to the accumulated relative phase, so that for certain values of $\theta - \phi$ the interference pattern is absent and the number state coefficients are given by a binomial distribution. On the other hand, if the initial state were a statistical mixture a binomial distribution would be expected for all values of the relative phase. 

\begin{figure}[t]
\begin{center}
\includegraphics[scale=0.35]{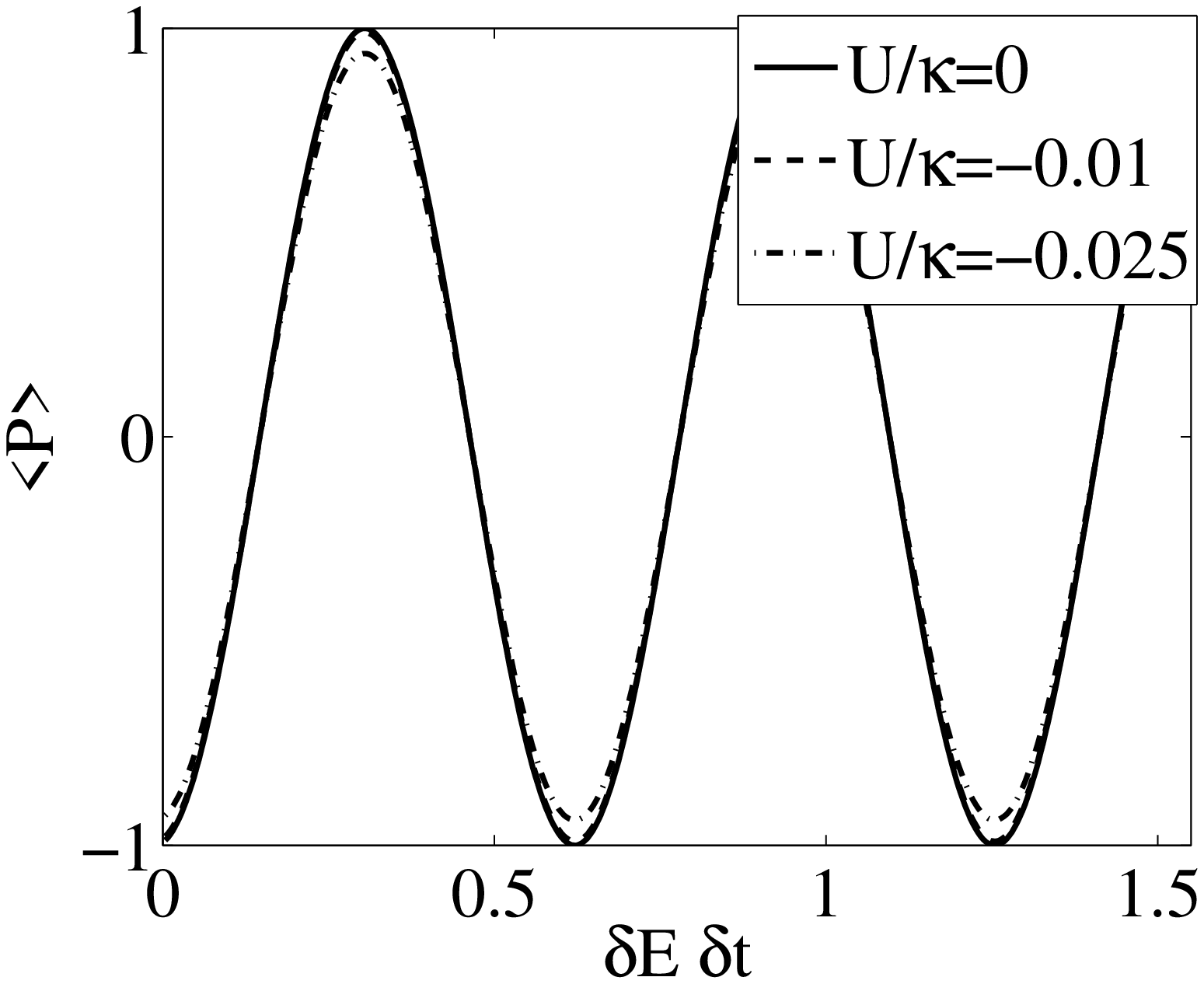}
\includegraphics[scale=0.35]{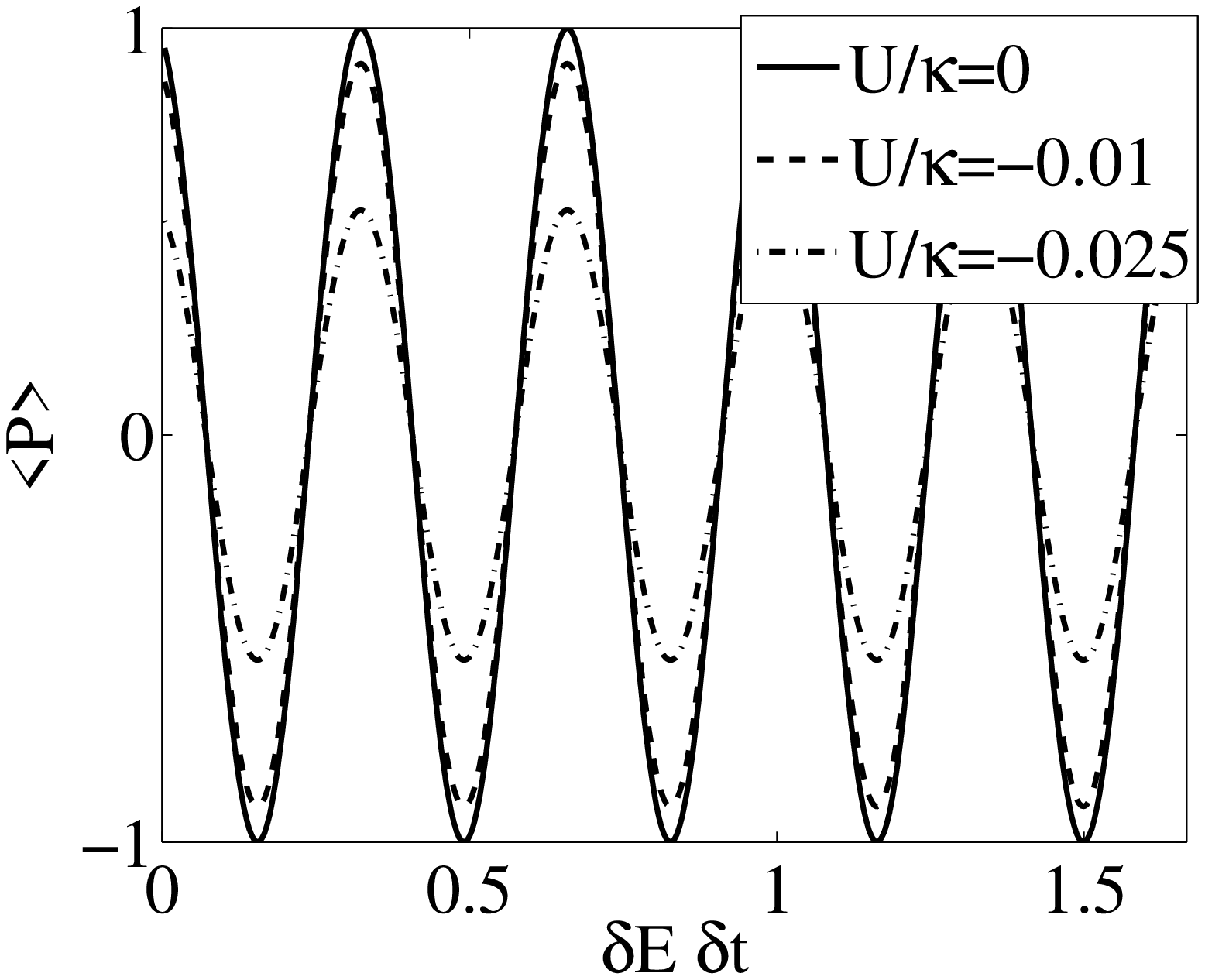}
\caption{Expectation value of parity after Ramsey simulation using ideal superposition state containing (a) 10, and (b) 20 atoms. Non-linear interactions reduce the amplitude of the parity oscillations. (Note that in (a) the $U/\kappa=-0.01$ line lies under the $U/\kappa=0$ line.)}
\label{parity}
\end{center}
\end{figure}

The interference can be seen most clearly in the parity, defined as 
\begin{equation}
  P = \sum_{n} (-1)^n \vert c_n \vert^2 ,
\end{equation}
which is the difference in probability that an even or odd number of atoms in measured in one of the modes at the end of the Ramsey experiment. The expectation value of the parity is an oscillatory function of the relative phase between the wells, again with a frequency proportional to the total atom number $N$, as seen in Fig.~\ref{parity} (a,b) . Measurements of the parity were considered in detail in Ref.~\cite{GerryCampos}, where the authors discuss methods for obtaining Heisenberg-limited phase resolution.

\subsection{Measurement difficulties}

Accurate measurement of the $N$th quadrature moment, or equivalently the atom parity after the Ramsey experiment, is experimentally very difficult. Measuring the interference pattern in the $c_n$ coefficients would require very accurate atom counting, as the relative phase value giving maximum fringe visibility for an even total atom number gives minimum fringe visibility if the total atom number is odd, and vice versa. Indeed, it is apparent that the loss of a single atom would completely destroy the coherence of the NOON state. Therefore, the required counting efficiency $\eta$ is such that $\eta^N \sim 1$, which would be challenging for large values of $N$. An alternative method of determining the parity has been suggested in Ref.~\cite{GerryCampos}. This method does not require extremely accurate atom counting; however it does involve a third condensate mode and precise control of the nonlinearity.

There are additional complications that may make this procedure difficult, even for moderate values of $N$. We now consider the effect of finite non-linear interactions during the Ramsey interference procedure. With an ideal superposition as the initial state, the non-linear interactions degrade the visibility of the interference pattern in the $c_n$ coefficients, as shown in Fig.~\ref{ramseycninterference} (b,c). The oscillations in the expectation value of parity are also reduced by the presence of non-linear interactions (see Fig.~\ref{parity}). For a given value of $U$, the visibility of the parity oscillations decreases as the number of atoms in the superposition increase. However, we note that the phase of the interference fringes is unaffected by the non-linearity, which is important for possible interferometric applications.

Finally, when the initial state is an ideal NOON state, we observed that including non-linear interactions can result in small fringes in the mean quadrature value even for $N \ge 2$. The maximum amplitude of these induced fringes is much less than the total atom number. These fringes are most noticeable for small atom numbers, as they have amplitudes of only one or two atoms regardless of the total atom number.

\section{Generating a superposition state}

\label{sec4}

There have been a number of proposals for generating superpositions of various kinds involving Bose-Einstein condensates. Many of these consider superpositions of two-component condensates. In this case, methods involving adiabatic manipulation~\cite{2compCirac1997}, and dynamical evolution (making use of the interplay between nonlinear interactions and tunneling, with a specific initial relative phase)~\cite{Dunningham,Zoller03,GordonSavage}, have been suggested to generate superpositions.  Superpositions of phase states have also been considered~\cite{phasesupn1,phasesupn2}. 

We are interested in superpositions of number states in a single component condensate in a double well potential. One proposal for generating superpositions of this kind involves using a Feshbach resonance to produce a sudden change in the interaction strength, initiating a dynamical evolution where a macroscopic superposition emerges periodically~\cite{HuangMoore}.

In Ref.~\cite{HuangMoore}, it was claimed that the adiabatic method was not feasible due to the infinitely long evolution time required because of the near degeneracy of the ground and first excited states in the strongly attractive regime. We find that with realistic parameters a superposition of two well-separated wave packets can be generated on a time scale of seconds using a smooth change in interaction strength. Although this evolution is not necessarily adiabatic it can generate a `mesoscopic' superposition state.

To estimate some suitable parameters for our effective Hamiltonian we have considered a small condensate of Rubidium atoms. Around a magnetic field strength of 155 G there is a Feshbach resonance for $^{85}$Rb atoms, allowing the s-wave scattering length to be tuned from at least 2000 $a_0$ to $-200~a_0$ ($a_0$ being the Bohr radius). Assuming a cigar-shaped condensate confined by trapping frequencies of 1 kHz in both tight directions and 100 Hz in the longitudinal direction, and using a Gaussian approximation for the condensate wavefunction, we estimate that the interaction parameter, $U$, given by equation (\ref{interactioneqn}) could range from approximately 30~s$^{-1}$ to $-3$~s$^{-1}$. We look at simulations which are a few seconds in length (note that condensate lifetimes of greater than 10 seconds are experimentally achievable~\cite{CornellWiemanlifetime}).

\begin{figure}[t]
\begin{center}
\includegraphics[width=0.34\columnwidth]{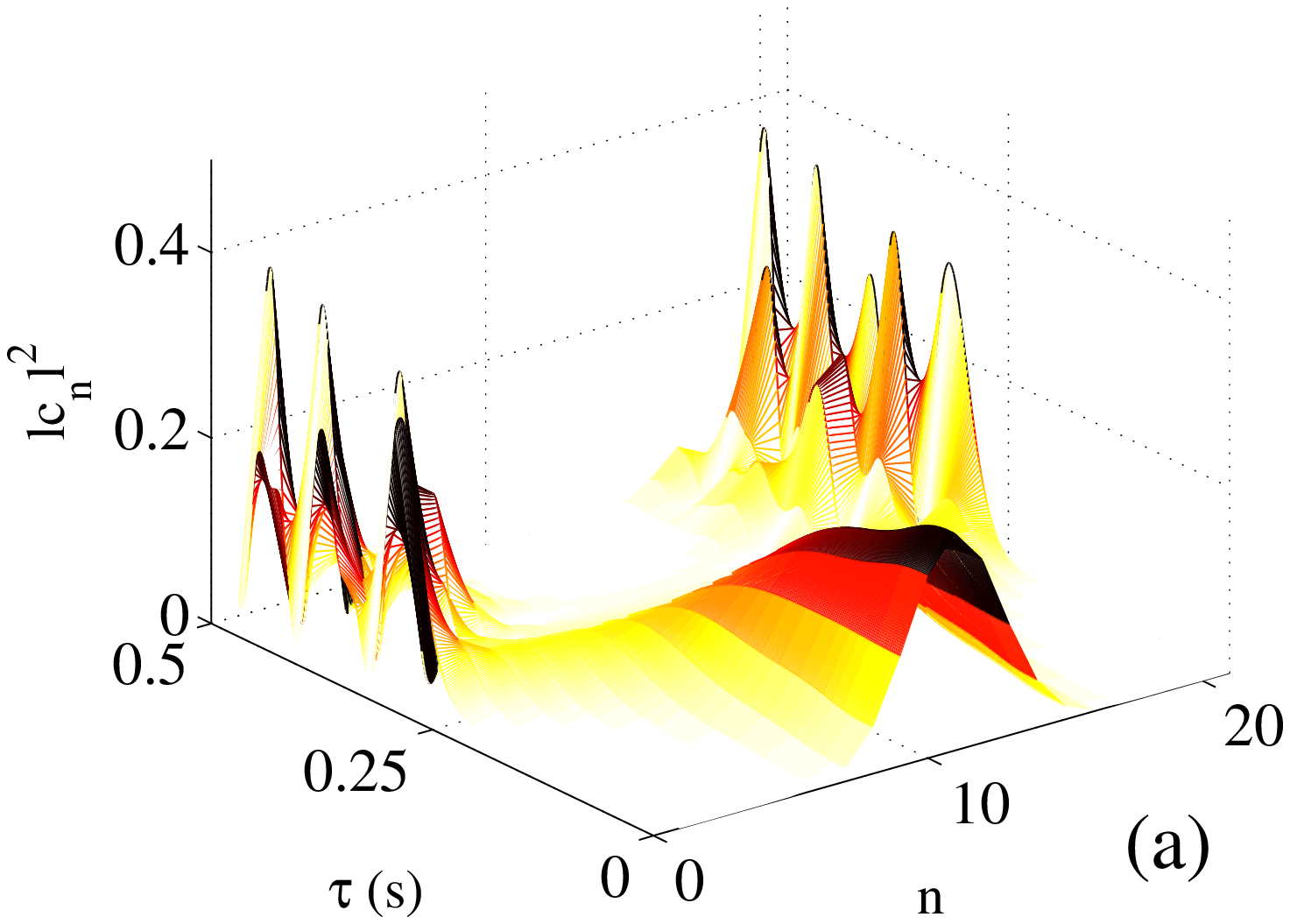}
\includegraphics[width=0.34\columnwidth]{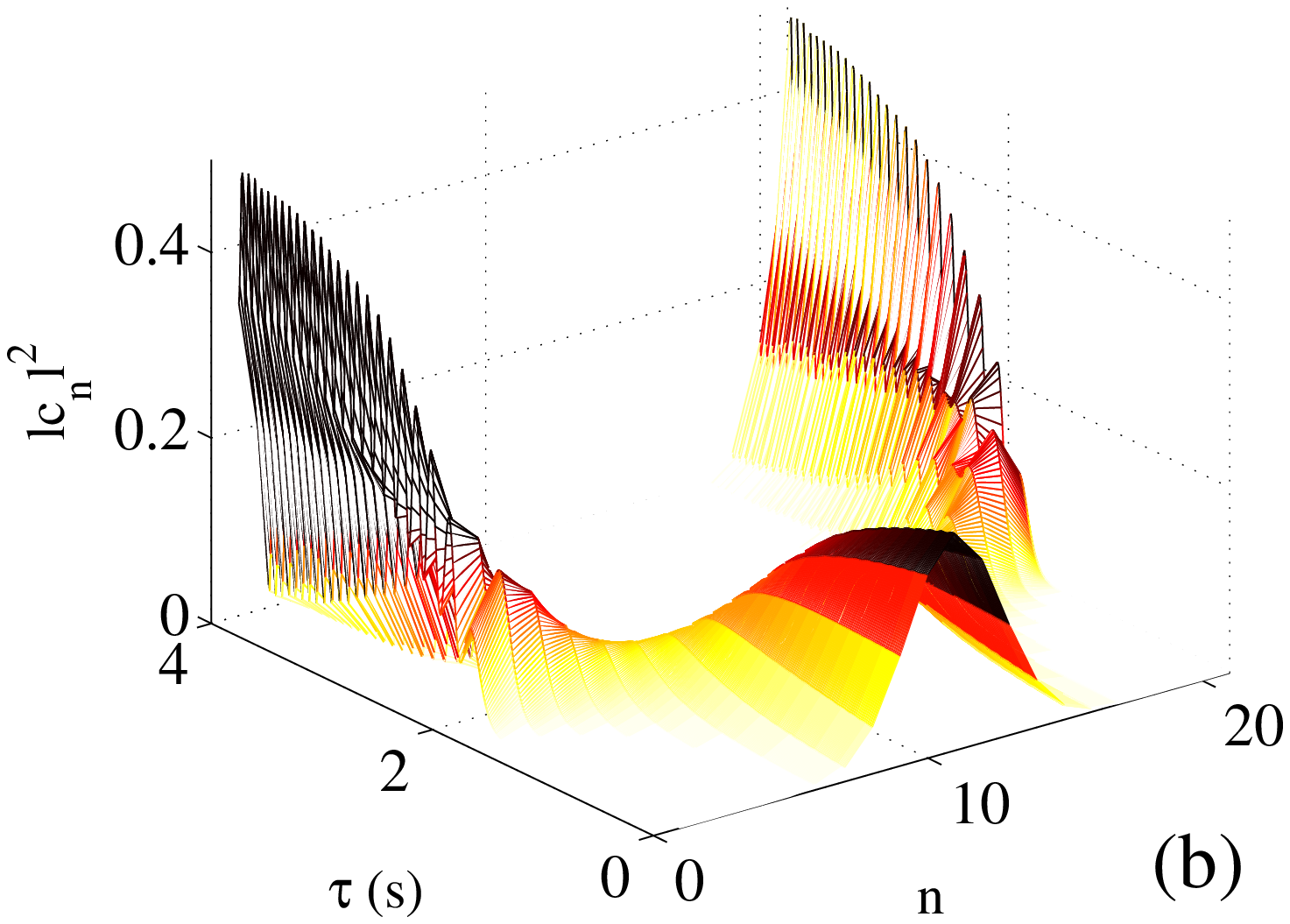}
\caption[Optional caption for list of figures]{Evolution of wave function into a mesoscopic superposition caused by a linear decrease in interaction strength from U=1~s$^{-1}$ to $-3$~s$^{-1}$ in (a) 0.5, and (b) 4 seconds. Initial wavefunction is the ground state at U=1~s$^{-1}$ and $\kappa=10$~s$^{-1}$.}
\label{linchangeU}
\end{center}
\end{figure}

\begin{figure}[t]
\begin{center}
\includegraphics[width=0.32\columnwidth]{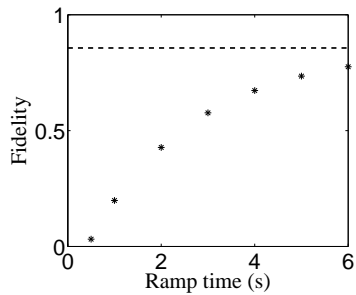}
\caption{Fidelity of final wave function compared to the ideal NOON state, for different ramp times (dots). Dashed line is fidelity of ground state $U=-3$~s$^{-1}$ and $\kappa=10$ compared to the ideal NOON state. A slower ramp time results in a final wave function closer to that of an ideal superposition state.}
\label{fidelity}
\end{center}
\end{figure}

Fig.~\ref{linchangeU} shows the wave function as the interaction strength is changed linearly from 1~s$^{-1}$ to $-3$~s$^{-1}$ over a timespan of 0.5 and 4 seconds. The initial state is the ground state at U=1~s$^{-1}$, and $\kappa=10$~s$^{-1}$. Clearly the slower change in interaction strength results, as per the adiabatic theorem, in a final state that is closer to the ideal macroscopic superposition. This can be seen in Fig.~\ref{fidelity}, which plots the fidelity of the evolving wave function, i.e. the overlap, $\vert \langle \phi_{NOON} \vert \phi(t)\rangle \vert^2$, of the ideal superposition state with the evolving wave function. The variance in the number difference between the wells of the final states in Fig.~\ref{linchangeU} are approximately 283 and 371, compared with 400 for an ideal superposition state containing 20 atoms. Once the interaction strength is held constant at $-3$~s$^{-1}$ the probability distributions $|c_n|^2$ do not change significantly with time. 

The parameters could be optimized further to obtain better superposition states. In general, better states are obtained when the final ratio $U/\kappa$ is large and negative. On the other hand, the timescale for the evolution to be adiabatic is inversely proportional to both $U$ and $\kappa$. The requirement of physically separated wells implies a relatively small value for $\kappa$, while the timespan of the experiment is limited by atomic loss. We have not attempted to optimize the process of generating a superposition state within these bounds, but rather concentrate on measurements aimed at distinguishing even an imperfect superposition from a statistical mixture.

\subsection{Intereference measurements of non-ideal superposition states}

\begin{figure}[t]
\begin{center}
\includegraphics[scale=0.32]{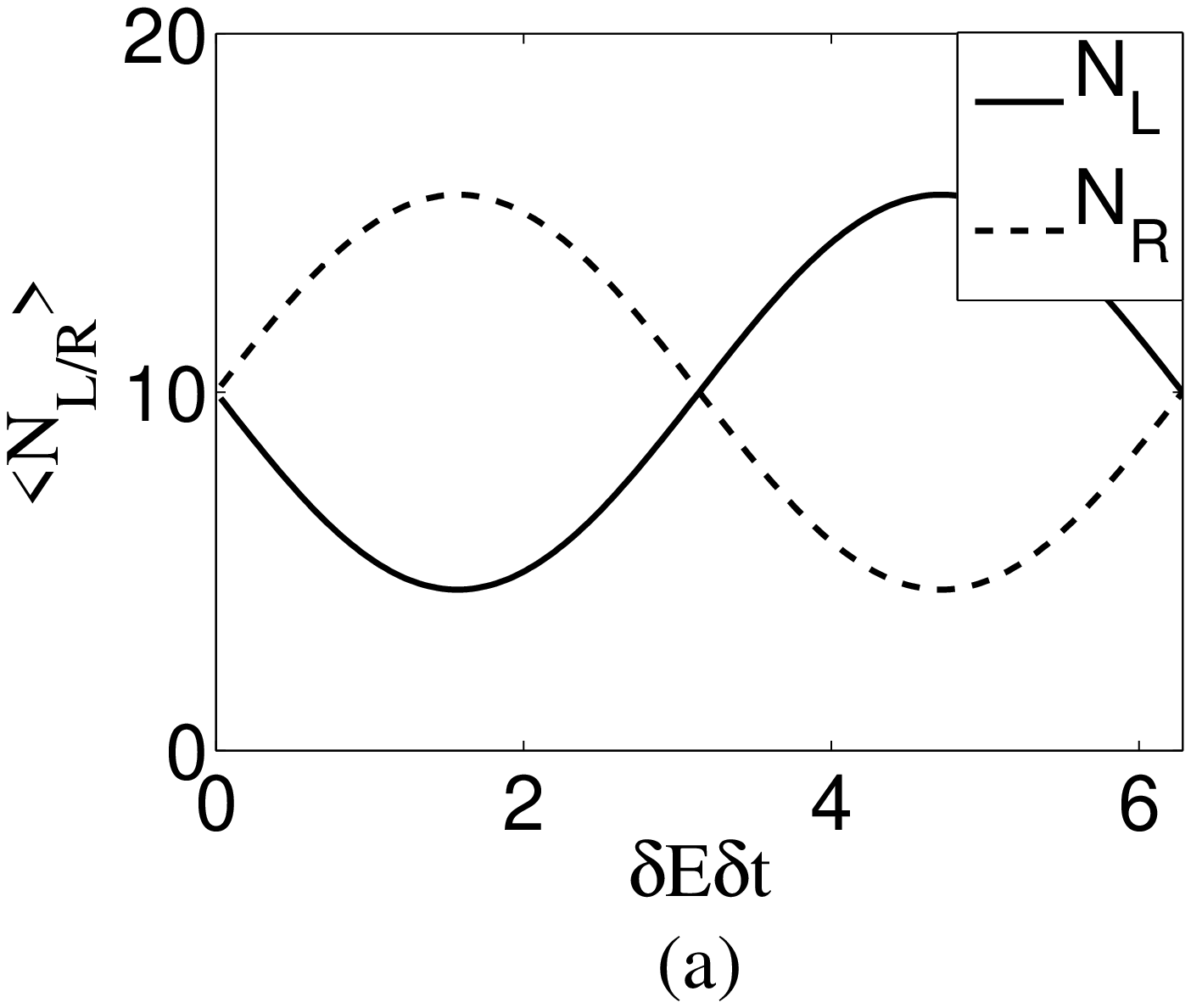}
\includegraphics[scale=0.32]{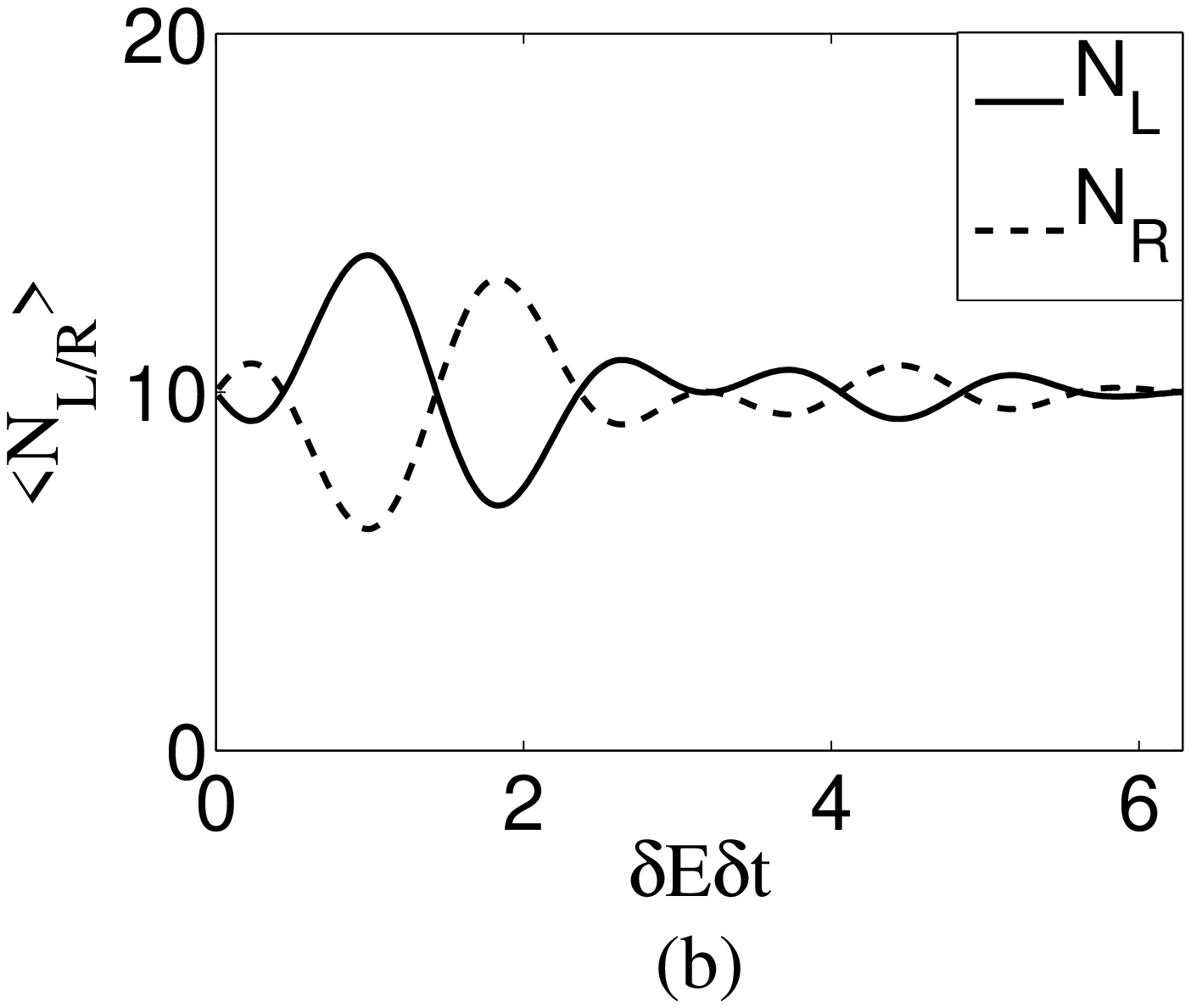}
\includegraphics[scale=0.32]{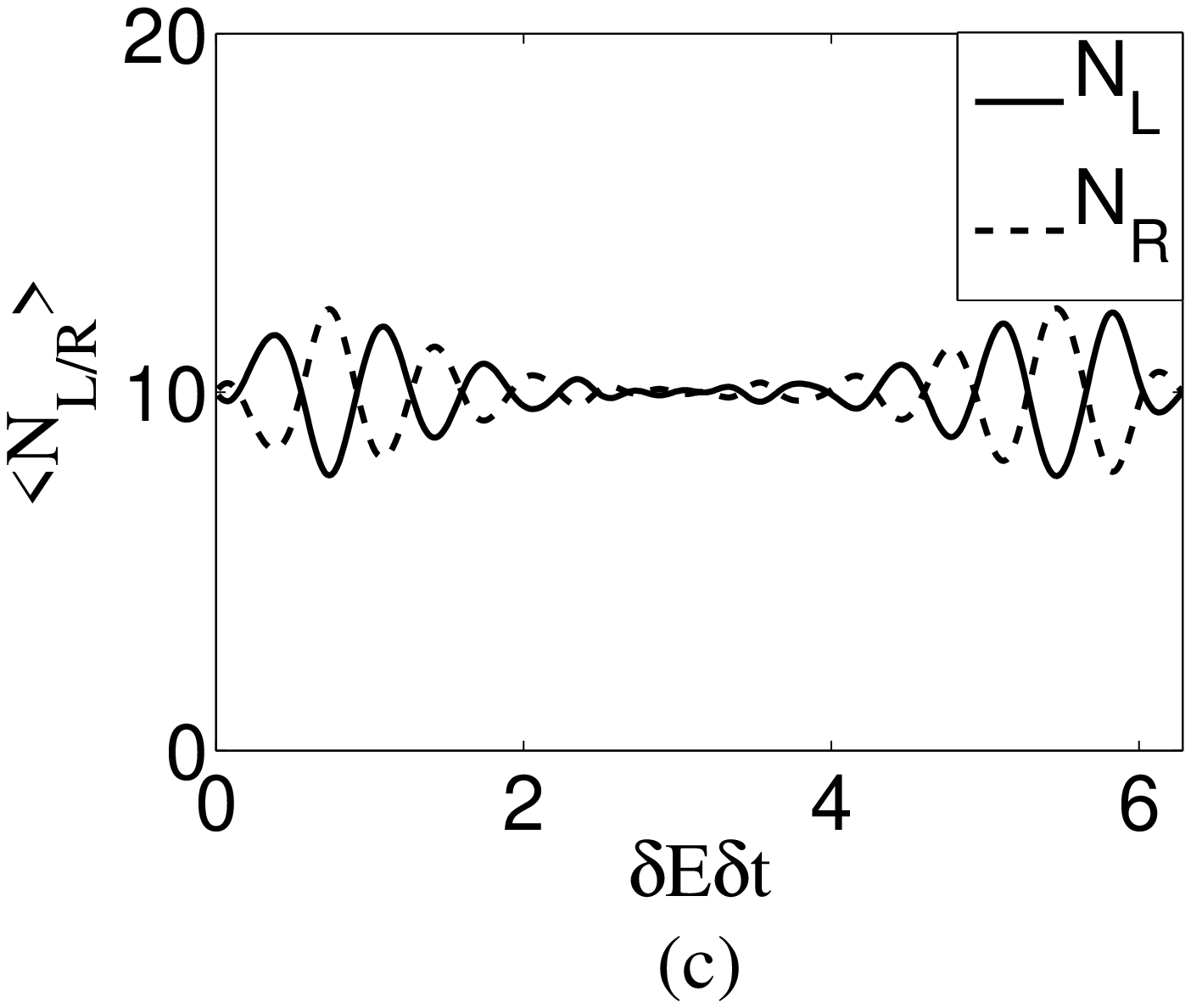}
\caption[Optional caption for list of figures]{Expectation value of atom number in each well after Ramsey simulation for a 20 atom condensate, for non-linear interactions being (a) zero, (b) $-0.1$~s$^{-1}$, and (c) $-0.25$~s$^{-1}$. Initial state was generated from the ground state at U=1 s$^{-1}$, $\kappa=10$ s$^{-1}$, and linearly changing the interaction strength to $-1$ s$^{-1}$ over 4 seconds.}
\label{ramseynonideal}
\end{center}
\end{figure}

\begin{figure}[t]
\begin{center}
\includegraphics[scale=0.32]{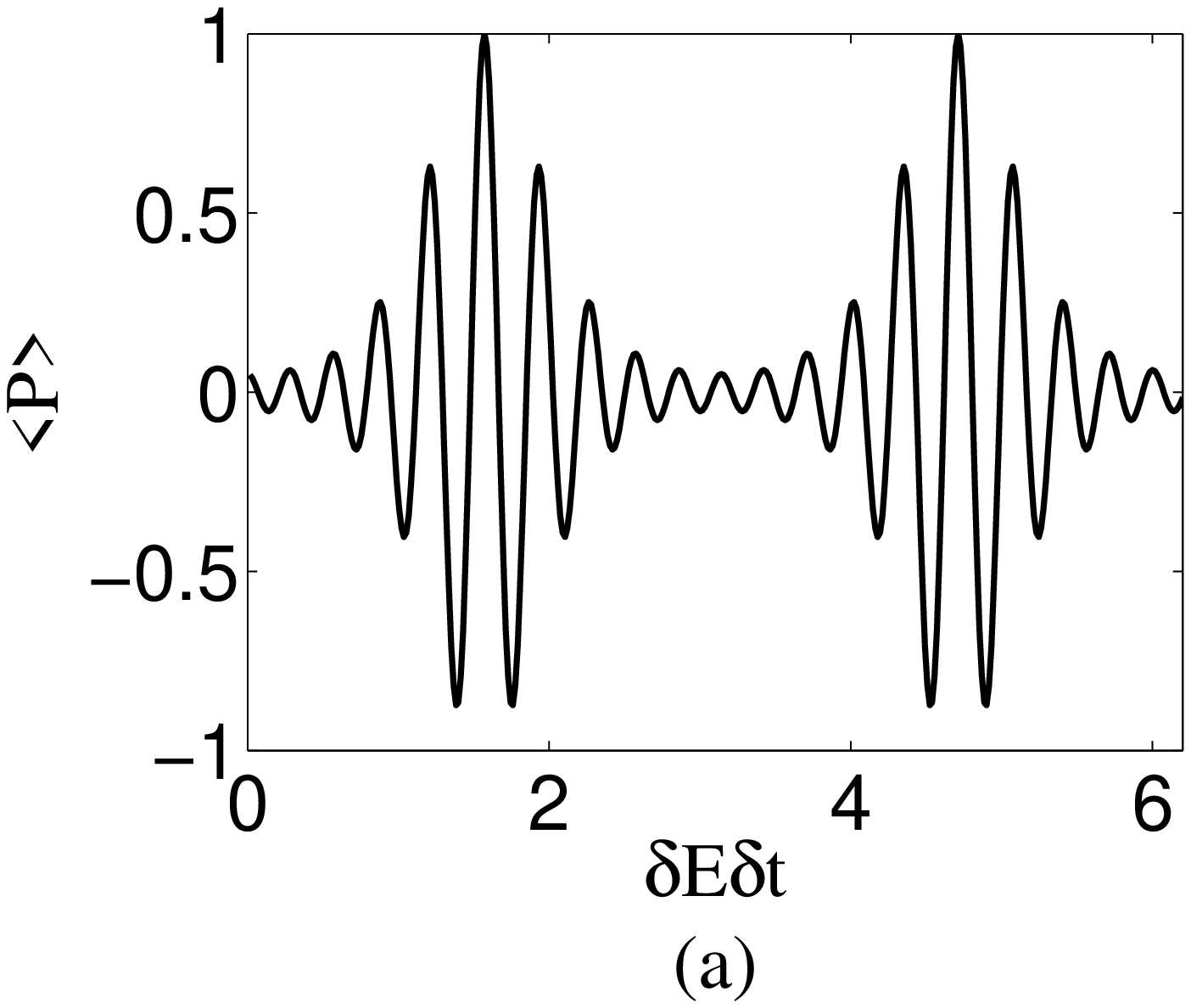}
\includegraphics[scale=0.32]{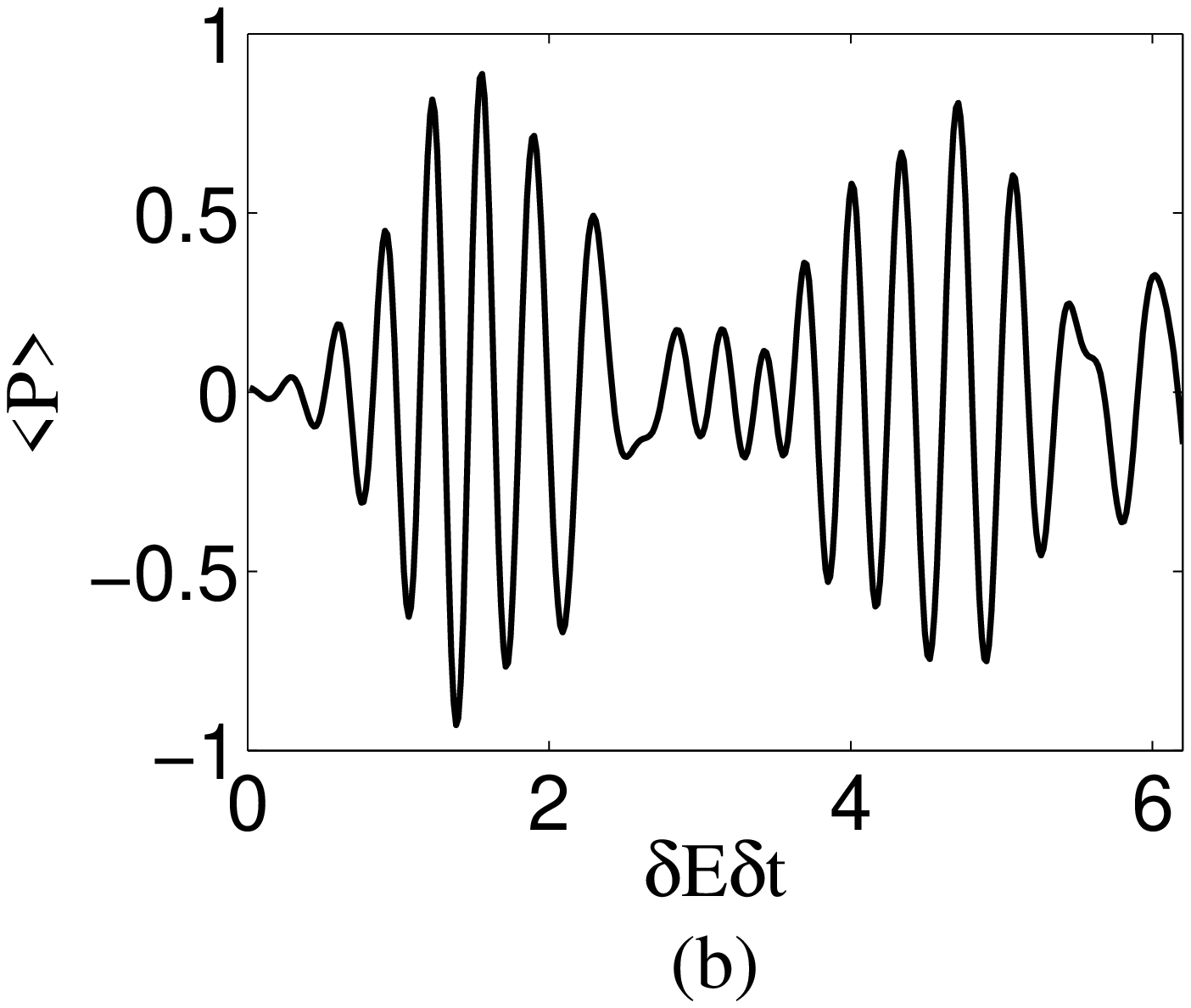}
\includegraphics[scale=0.32]{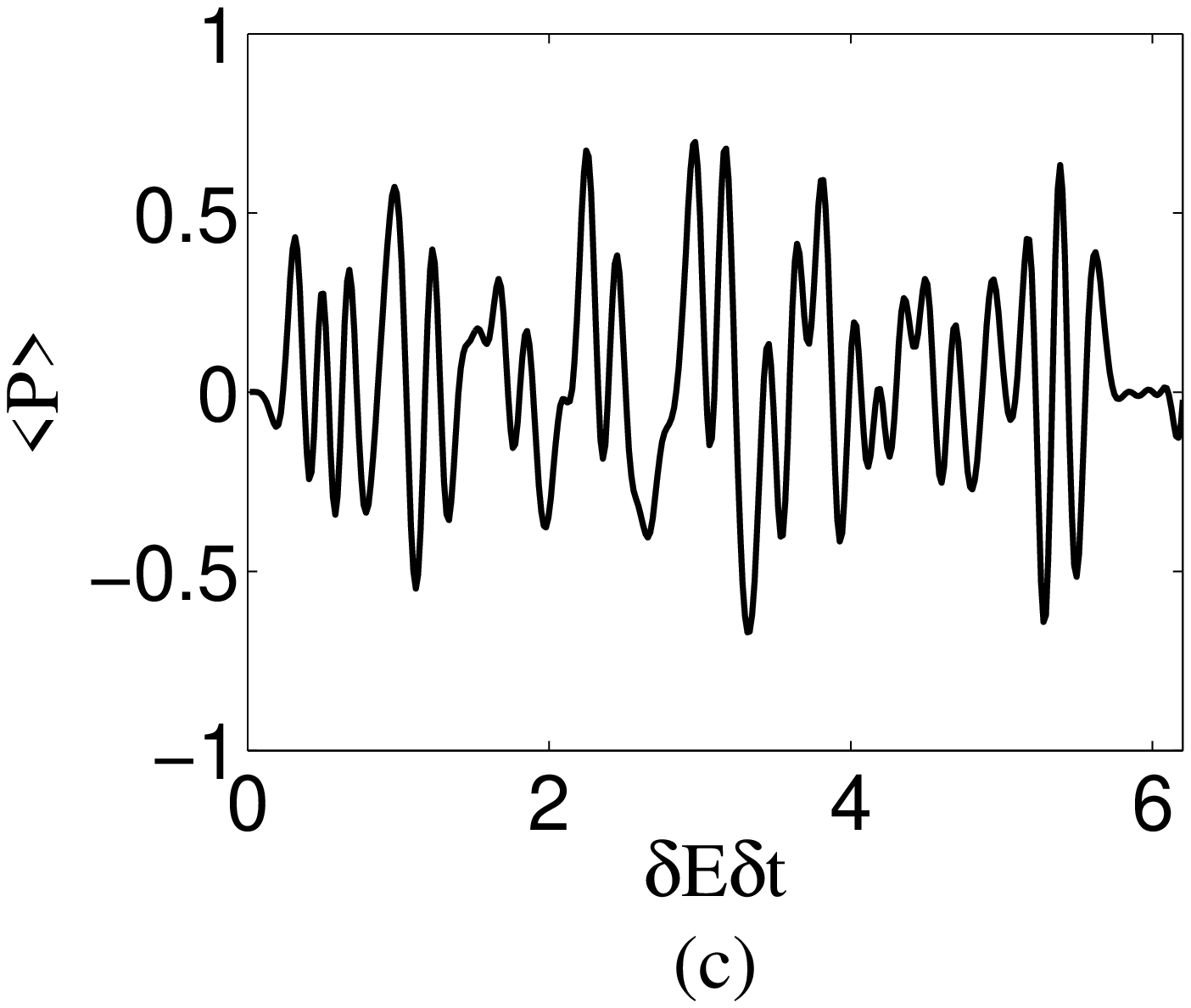}
\caption[Optional caption for list of figures]{Expectation value of parity after Ramsey simulation of a 20 atom condensate, for non-linear interactions being (a) zero, (b) $-0.1$~s${}^{-1}$, and (c) $-0.25$~s${}^{-1}$. Initial state was generated from the ground state at U=1 s${}^{-1}$, $\kappa=10$ s${}^{-1}$, and linearly changing the interaction strength to $-1$ s${}^{-1}$ over 4 seconds.
}
\label{paritynonideal}
\end{center}
\end{figure}

We now perform the Ramsey procedure followed by parity measurements in order to detect coherence in the non-ideal superposition states generated above. Figs.~\ref{ramseynonideal} and \ref{paritynonideal} show typical fringes in the number difference and parity for an initial non-ideal superposition state of a 20 atom condensate. In the parity, we see clear high frequency components in the interference fringes that correspond to the off-diagonal coherence between states of large atom number difference. However, as the state is not perfect, other frequency components are present, resulting in beating and a low frequency envelope.

In the Appendix, we formalise the relationship between the parity frequency components and coherence and show that
\begin{equation}
  P = \sum_{n,m} B_{n,m} e^{im\phi} \langle N-n,n| \hat{\rho} | N-n-m, n+m \rangle + \mathrm{H.c.} \label{frequencies}
\end{equation}
for some real numbers $B_{n,m}$. The component with angular frequency $m$ corresponds to coherence between elements separated by $m$ atoms in Fock space. In fact, the highest frequency component with angular frequency $N$ is proportional only to $\langle N,0| \hat{\rho} |0,N\rangle$, and is the only component observed in the results for an ideal superposition (Fig.~\ref{parity}). The presence of the same high frequency component in Fig.~\ref{paritynonideal} (a) is an unambiguous demonstration that the generated state contains coherence and is \emph{not} a statistical mixture.

With the introduction of non-linear interactions during the Ramsey interference, the amplitude and frequency of the parity fringes become less regular and are no longer periodic over $\theta - \phi$ modulo $2\pi$. It would not be easy to use the results when $U$ is large as an indicator of the presence of a superposition due to their irregularity, and the fact that high frequency components cannot be said to indicate NOON-type coherence.

\section{Conclusions}

\label{sec5}

Our simulations indicate that a quantum superposition of a small Bose-Einstein condensate may be generated in a reasonably short time by a smooth change in the atomic interaction strength from repulsive to attractive. However, we find that verifying that a superposition has been generated may not be straightforward, even in the ideal case of a perfect `NOON' state superposition and no decoherence. We have considered a Ramsey-type interference experiment as a method to measure the quadrature operator and parity thereof of a double well wave function. 

For a NOON superposition, only the highest-order quadrature moment (and not the mean) is sensitive to the accumulated relative phase and is therefore useful in distinguishing between a statistical mixture and a coherent superposition. Parity measurements after a Ramsey-type experiment on an ideal superposition state show a dependence on the relative phase that could in principle be used to verify the presence of the superposition. A statistical mixture would not be expected to show any dependence on the relative phase if parity measurements were performed. The presence of non-linear interactions during the Ramsey experiment degrades the amplitude of the parity oscillations.

Measurements of a non-ideal superposition state also show a high-frequency dependence on the relative phase --- a smoking gun indication of NOON-type coherence. However, we observed that the imperfect state displays additional frequencies resulting in a beating pattern. For some choices of phase angle, the visibility is close to one while the period decreases to $2\pi/N$, allowing it to remain useful for Heisenberg-limited phase measurements. Similarly as for the ideal state, non-linear interactions degrade the amplitude and regularity of the parity oscillations, however for small non-linearity, the high frequency component could still be used to verify that a mesoscopic superposition has been generated.

There are many factors that may make these measurements difficult in practice. For example, phase diffusion due to non-linear interactions dramatically reduces the visibility of the parity oscillations. Other effects not considered here would also reduce the fringe visibility, such as 3-body loss, and other sources of decoherence. Highly accurate atom counting is required to observe the interference patterns in the $c_n$ coefficients, and the associated parity oscillations as a function of accumulated phase, unless an alternative measurement scheme could be realised.

\ack

The authors wish to acknowledge support from the Australian Research Council.

\appendix

\section*{Appendix}

For a state with precisely $N$ atoms, the (discrete) quadrature operator can take $N+1$ values. Therefore, it is possible to express any linear combination of the measurement probabilities (i.e.~the $\vert c_n \vert^2$ after the Ramsey interference) as a linear sum of the first $N$ quadrature moments. Specifically, the parity can be expressed this way, i.e.
\begin{eqnarray}
 P(\theta) & = & \sum_{n=0}^N A_n \langle \hat{\mathcal{X}}_{\theta}^n \rangle \nonumber \\
 & = & \sum_{n=0}^N A_n \langle \left(e^{i\theta} \hat{a}_L^{\dag} \hat{a}_R + e^{-i\theta} \hat{a}_L \hat{a}_R^{\dag} \right)^n \rangle ,
\end{eqnarray}
where $\{A_n\}$ are real numbers. Expanding the power results in
\begin{equation}
 P(\theta) = \sum_{n=0}^N \sum_{m=0}^n A_n \, ^m\mathrm{C}_n e^{(n-2m)\theta} \langle :\hat{a}_L^{\dag \,n-m } \hat{a}_L^m \hat{a}_R^{\dag \, m} \hat{a}_R^{n-m}:_{\mathrm{sym}} \rangle ,
\end{equation}
where $: \hat{x} :_{\mathrm{sym}}$ represents the symmetric ordering of $\hat{x}$. 

The operator $:\hat{a}_L^{\dag \,n-m } \hat{a}_L^m \hat{a}_R^{\dag \, m} \hat{a}_R^{n-m}:_{\mathrm{sym}}$ transfers $n - 2m = M$ atoms from the right well to the left, and therefore we can say
\begin{equation}
  \langle :\hat{a}_L^{\dag \,n-m } \hat{a}_L^m \hat{a}_R^{\dag \, m} \hat{a}_R^{n-m}:_{\mathrm{sym}}\rangle = \sum_n D_n \langle N-n,n| \hat{\rho} |N-n-m,n+m\rangle,
\end{equation}
where $D_n$ are appropriately chosen constants. 

Putting everything together, we finally arrive at Eq.~(\ref{frequencies}), where the the constants $B_{n,m}$ can be obtained from the above.


\section*{References}

\begin{thebibliography}{99} 
%*******************************************


\bibitem{schrodingerscat}{Schr\"{o}dinger E, Translation by Trimmer J D {\it{Proceedings of American Philosophical Society}} {\bf 124} 323}

\bibitem{beryllium}{Leibfried D, Knill E, Seidelin E, Britton J, Blakestad R B, Chiaverini J, Hume D B, Itano W M, Jost J D, Langer C, Ozeri R, Reichle R and Wineland D J, Nature 2005 {\it{Nature}} {\bf 438} 639}

\bibitem{SQUIDs}{Friedman J R, Patel V, Chen W, Tolpygo S K and Lukens J E 2000 {\it{Nature}} {\bf 406} 43}

\bibitem{Reid}{Cavalcanti E and Reid M D 2008 \PRA {\bf 77} 062108}

\bibitem{2compCirac1997} {Cirac J I, Lewenstein M, M\o{}lmer K and Zoller P 1998 \PRA {\bf 57} 1208}

\bibitem{Dunningham} {Dunningham J A, and Burnett K 2001 {\it{Journal of Modern Optics}} {\bf 48} 1837}

\bibitem{Zoller03}{Micheli A, Jaksch D, Cirac J I and Zoller P 2003 \PRA {\bf 67} 013607}

\bibitem{GordonSavage}{Gordon D and Savage C M 1999 \PRA {\bf 59} 4623}

\bibitem{phasesupn1}{Piazza F, Pezz\'{e} L and Smerzi A 2008 \PRA {\bf 78} 051601}

\bibitem{phasesupn2}{Dunningham J A, Burnett K, Roth R and Phillips W D 2006 {\it New Journal of Physics} {\bf 8} 182}

\bibitem{dwsqueezing}{Est\`{e}ve J, Gross C, Weller A, Giovanazzi S and Oberthaler M K 2008 {\it Nature} {\bf 455}}

\bibitem{dwinterferometrychip}{Schumm T, Hofferberth S, Andersson L M, Wildermuth W, Groth S, Bar-Joseph I, Schmiedmayer J and Kr\"{u}ger P 2005 {\it Nature} {\bf 1} 57}

\bibitem{TindleWalls}{Tindle C T and Walls D F 1972 \JPA {\bf 4} 534}

\bibitem{ITprop}{Balbuena P B and Seminario J M 1999 {\emph{Molecular Dynamics: From Classical to Quantum Methods}} (Amsterdam: Elsevier Science) p~149}

\bibitem{HuangMoore}{Huang Y P and Moore M G 2006 \PRA {\bf 73} 023606}

\bibitem{CornellWiemanlifetime}{Stamper-Kurn D M, Andrews M R, Chikkatur A P, Inouye S, Miesner H J, Stenger J and Ketterle W 1998 \PRL {\bf 80} 2027}

\bibitem{EPRquad}{Kheruntsyan K V, Olsen M K and Drummond P D 2005 \PRL {\bf 95} 150405}

\bibitem{Ferris2008}{Ferris A J, Olsen M K, Cavalcanti E G and Davis M J 2008 \PRA {\bf 78} 060104(R)}

\bibitem{Ferris2009}{Ferris A J, Olsen M K and Davis M J 2009 \PRA {\bf 79} 043634}

\bibitem{Ramsey}{Ramsey N F 1980 \emph{Phys. Today} {\bf 33} 7}

\bibitem{Corney}{Corney J F, Milburn G J and Zhang W 1998 \PRA {\bf 59} 4630}

\bibitem{PezzeSmerzi}{Pezz\'{e} L, Smerzie A, Berman G P, Bishop A R and Collins L A 2006 \PRA {\bf 74} 033610}

\bibitem{Higgins}{Higgins B L, Berry D W, Bartlett S D, Wiseman H M, and Pryde G J 2007 \emph{Nature} {\bf 450} 393}

\bibitem{GerryCampos}{Gerry C C and Campos R A 2003 \PRA {\bf 68} 025602}

\end{thebibliography}
\end{document}